\documentclass[journal]{IEEEtran}

\usepackage{graphicx}
\usepackage{epstopdf}

\usepackage{subfigure}
\usepackage{booktabs}
\usepackage{caption}
\usepackage{color}
\usepackage{bm}
\usepackage{CJK}
\usepackage{indentfirst}
\usepackage{amsmath}
\usepackage{amssymb}
\usepackage{float}
\usepackage{multirow}
\usepackage{algorithm}
\usepackage{algpseudocode}
\usepackage{amsmath}
\usepackage{flushend}
\usepackage{geometry}

\begin{document}
\title{{ Channel Customization for Low-Complexity CSI Acquisition in Multi-RIS-Assisted MIMO Systems}}
\author{\IEEEauthorblockN{Weicong~Chen, \textit{Member, IEEE}, Yu~Han, \textit{Member, IEEE}, Chao-Kai~Wen, \textit{Fellow, IEEE}, Xiao~Li, \textit{Member, IEEE}, and~Shi~Jin, \textit{Fellow, IEEE}}\\
\thanks{{Weicong Chen, Yu Han, Xiao Li, and Shi Jin are with the National Mobile Communications Research Laboratory, Southeast University,
Nanjing, 210096, P. R. China (e-mail: cwc@seu.edu.cn; hanyu@seu.edu.cn; li\_xiao@seu.edu.cn; jinshi@seu.edu.cn).} }
\thanks{Chao-Kai Wen is with the Institute of Communications Engineering, National Sun Yat-sen University, Kaohsiung 80424, Taiwan. (e-mail: chaokai.wen@mail.nsysu.edu.tw).}
}
\maketitle\thispagestyle{empty}

\begin{abstract}
The deployment of multiple reconfigurable intelligent surfaces (RISs) enhances the propagation environment by improving channel quality, but it also complicates channel estimation. Following the conventional wireless communication system design, which involves full channel state information (CSI) acquisition followed by RIS configuration, can reduce transmission efficiency due to substantial pilot overhead  and computational complexity. This study introduces an innovative approach that integrates CSI acquisition and RIS configuration, leveraging the channel-altering capabilities of the RIS to reduce both the overhead and complexity of CSI acquisition. The focus is on multi-RIS-assisted systems, featuring both direct and reflected propagation paths. By applying a fast-varying reflection sequence during RIS configuration for channel training, the complex problem of channel estimation is decomposed into simpler, independent tasks. These fast-varying reflections effectively isolate transmit signals from different paths, streamlining the CSI acquisition process for both uplink and downlink communications with reduced complexity. In uplink scenarios, a positioning-based algorithm derives partial CSI, informing the adjustment of RIS parameters to create a sparse reflection channel, enabling precise reconstruction of the uplink channel. Downlink communication benefits from this strategically tailored reflection channel, allowing effective CSI acquisition with fewer pilot signals. Simulation results highlight the proposed methodology's ability to accurately reconstruct the reflection channel with minimal impact on the normalized mean square error while simultaneously enhancing spectral efficiency.
\end{abstract}



\section{Introduction}
\IEEEPARstart{5G}{technology} has significantly expanded the capabilities and applications of wireless communications, driving the demand for superior service experiences and creating opportunities for targeted industry advancements. This progress highlights the critical importance of understanding and managing the wireless channel state to develop sophisticated transmission strategies. Historically, the focus of wireless system development has been on enhancing transmitter and receiver technologies to counteract the effects of wireless fading. Substantial advancements have been made in areas such as multiple antenna systems \cite{e-MIMO}, channel coding \cite{e-coding}, waveform design \cite{e-waveform}, and multiple access methods \cite{e-access}. Nevertheless, the challenge of signal fading has often relegated the wireless channel to a secondary concern, typically addressed through relay technologies \cite{relay}. While these technologies improve transmission quality, they also introduce additional noise and errors during signal processing. 

Recent advancements have shifted the focus toward the joint optimization of transmitters, channels, and receivers, with reconfigurable intelligent surfaces (RISs) emerging as a key technology.  Originating from metamaterials \cite{meta-Cui}, RISs have a simple hardware structure and a quasi-passive, lightweight nature, making them cost-effective, low-power, and easy to deploy on a large scale. They offer unprecedented control over electromagnetic waves, enabling the creation of ``smart'' wireless channels \cite{smart-radio} that can dynamically adapt to communication needs. The theoretical and practical benefits of RISs for enhancing wireless communication systems have been extensively documented. Examples include their use in resource allocation to improve energy efficiency \cite{app-19-Huang}, UAV communications for optimized data rates \cite{app-20-UAV}, and the design of specific transmission channels to meet diverse communication requirements \cite{CC-JSAC}. Further studies have explored the integration of RISs into cell-free networks \cite{app-22-cell-free}, satellite-terrestrial relay networks \cite{app-22-ZLin}, and millimeter-wave (mmWave) systems \cite{app-23-mmwave}, demonstrating their potential to significantly enhance network capacity, signal quality for obstructed users, and overall system performance in integrated sensing and communication systems \cite{app-22-ISAC}. 

The deployment of multiple RISs can effectively enrich the propagation environment, improving channel rank and ensuring uninterrupted connectivity within a target area \cite{Multi-RIS-1}--\cite{Multi-RIS-3}. However, the successful deployment of RISs depends heavily on the accurate acquisition of channel state information (CSI). This challenge is particularly pronounced in RIS-assisted systems, where conventional digital signal processing capabilities are limited due to the typically large number of passive elements that constitute the RIS \cite{CE-22-AL}. Various strategies have been proposed to address this issue, including exploiting channel correlations \cite{CE-20-Zwang}, employing atomic norm minimization for mmWave communications \cite{CE-21-JHe}, and developing joint localization and channel reconstruction schemes for near-field communications \cite{CE-22-YHan}. Methods such as anchor-assisted channel estimation \cite{CE-22-XGuan}, deep learning-based channel recovery \cite{CE-23-Wshen}, and Bayesian techniques for multi-RIS systems \cite{CE-20-GC} showcase the diverse approaches being explored to ensure accurate CSI under varying system conditions.
 
The implications of CSI estimation errors on system design have been a critical area of research, with studies examining robust beamforming \cite{IP-20-JZhang}, cooperative beamforming for double-RIS \cite{IP-21-CYou}, rate maximization in cognitive radio systems \cite{IP-21-JYuan}, and the impact on system performance metrics such as outage probability, bit error rate, and capacity \cite{IP-22-PYang}. Despite the challenges posed by imperfect CSI, research such as \cite{IP-22-KZ} and \cite{IP-24-DP} demonstrates the resilience of RIS-assisted systems, showing that with strategic design adjustments, these systems can maintain high-performance levels, underscoring the transformative potential of RIS technology.

Building upon the established research on RIS-assisted communications, this study introduces a paradigm shift by proposing a \emph{joint} framework for channel estimation and RIS configuration. This framework aims to streamline the CSI acquisition process for both uplink and downlink transmissions. By leveraging the RIS's ability to modify wireless channel characteristics, we propose a novel approach that deviates from the traditional sequential process of channel estimation followed by RIS configuration. Our approach involves extracting partial CSI during the uplink phase to inform the design of RIS configurations, thereby creating a sparse reflection channel. This channel simplification aids in reconstructing the uplink channel and estimating the downlink CSI. The key contributions of this research are as follows: 
\begin{itemize}
\item \emph{A Channel Customization Framework for CSI Acquisition in Multi-RIS-Assisted Systems:} 
    This contribution presents a novel scheme for channel separation by applying distinct, rapidly varying reflection phases across RIS elements. The scheme effectively separates transmit signals from direct and reflected paths, facilitating channel estimation in environments with complex multi-path reflections. By using partial CSI to configure RIS elements, we achieve channel sparsification, reducing the complexity of the estimation process in scenarios characterized by rich scattering. 

\item \emph{A Low-Complexity Algorithm for Extracting Path Parameters for Channel Sparsification:} 
    Unlike traditional methods that treat channel estimation and RIS configuration as separate processes, this study integrates the RIS's channel-altering capabilities directly into the channel estimation phase. We propose a positioning-based algorithm for the simultaneous extraction of line-of-sight (LoS) path parameters, which are then used to fine-tune RIS configurations. This strategy enhances the predominant LoS paths cascaded by RISs, enabling the approximation of the reflection channel through these paths and eliminating the need for exhaustive estimation of additional path parameters. 

\item \emph{A Low-Overhead Estimation Scheme for Downlink Channel Reconstruction:} 
    By employing the joint uplink estimation and RIS configuration approach, we tailor a sparse reflection channel that simplifies the downlink channel estimation process, aligning it with scenarios devoid of RIS. The adoption of rapidly changing phases in the RIS configurations reduces the downlink estimation task to a set of singular path detection problems. This approach facilitates user equipment (UE) self-positioning using fixed RIS positions and the estimated LoS paths in the UE-RIS channel. 

\item \emph{Comprehensive Simulations for the Proposed CSI Acquisition:} 
    Extensive simulations demonstrate the effectiveness of the proposed fast-varying RIS phase approach in segregating training signals from varying links. With the proposed channel customization method, RIS-enhanced paths become predominant. The simulations validate the precision of path parameter extraction and UE positioning under the customized sparse channel conditions. Although the novel CSI acquisition method introduces a marginal increase in normalized mean square error (NMSE) compared to traditional full channel estimation, it maintains comparable spectral efficiency (SE). 
\end{itemize}

The remainder of this paper is structured as follows: Section \ref{sec:2} outlines the system model foundational to this research. Section \ref{sec:3} delves into channel separation and the exhaustive extraction of channel parameters. Section \ref{sec:4} introduces the proposed methodology for low-complexity CSI acquisition through channel customization. Section \ref{sec:5} presents a comprehensive analysis of the numerical results validating our approach. Finally, Section \ref{sec:6} concludes the study, summarizing its findings and implications.

\emph{Notations}: Lowercase and uppercase letters denote vectors and matrices, respectively. The transpose, conjugate, and conjugate-transpose operations are represented by superscripts $(\cdot)^T$, $(\cdot)^*$, and $(\cdot)^H$, respectively. The function ${\rm tr}(\cdot)$ calculates the trace of a matrix. The notation ${\mathbb C}^{a\times b}$ represents the set of $a\times b$ complex matrices. The symbols $|\cdot|$ and $|\cdot|$ denote the absolute value and Euclidean norm, respectively. The Kronecker, Khatri-Rao, and Hadamard products are denoted by $\otimes$, $\oplus$, and $\odot$, respectively. The notation ${\mathbb E}{\cdot}$ represents the statistical expectation. The subscripts and/or superscripts $\rm u$, $\rm b$, $\rm r$, $\rm A$, and $\rm D$ of a variable indicate that the variable is related to the UE, the BS, the RIS, the angle of arrival (AoA), and the angle of departure (AoD), respectively. Table I provides the definition of primary parameters in alphabetical order.

\begin{table}[h]
	\caption{Definition of primary parameters}
	\centering
	\label{Tab:1}
	\textcolor{black}{\begin{tabular}{ll}
		\toprule
		{\bf Parameter} & {\bf Description}\\
		\midrule
		${\bf a}_{{\rm b}/{{\rm r},k}/{\rm u}}(\cdot)$ & array response vector for the BS/RIS$_k$/UE\\
		${\bf B}_{k,j}$&normalized channel component of path $j$ in ${{\bf{H}}_{{{\rm rb}},k}}$ \\
		$d_0$/$d_{k}^{{\rm ur}}$/$d_{k}^{{\rm rb}}$ & distance for the direct/UE--RIS$_k$/RIS$_k$--BS channel\\
		${\bf e}_{{\rm b}}$/	${\bf e}_{{k}}$/	${\bf e}_{{\rm u}}$ & Position of the BS/RIS$_k$/UE\\
		$f_{k,v}$/${\bf f}_{k}$/${\bf F}$& fast-varying reflection coefficient/vector/matrix \\
		$g_{k,l}^{{\rm ur}}$/$g_{k,l}^{{\rm rb}}$& effective gain of path $l$ in ${{\bf{H}}_{{{\rm ur}},k}}$ / ${{\bf{H}}_{{{\rm rb}},k}}$ \\
		${{\bf{H}}_{{{\rm ur}},k}}$ / ${{\bf{H}}_{{{\rm rb}},k}}$  &the  UE--RIS$_k$/RIS$_k$--BS channel \\
		$L_k^{{\rm ur}}$/$L_k^{{\rm rb}}$&number of NLoS paths  in ${{\bf{H}}_{{{\rm ur}},k}}$ / ${{\bf{H}}_{{{\rm rb}},k}}$ \\
		${\bf U}_{k,j}$&normalized channel component of path $j$ in ${{\bf{H}}_{{{\rm ur}},k}}$ \\
		$M_k$& number of elements at the RIS$_k$\\
		$N_{\rm u}$/$N_{\rm b}$ & number of antennas at the UE/BS\\
		$p_{i}$/${\bf P}$&power allocation coefficient/matrix \\
		${\bf r}_{s,p,v}^{\rm b}$/${\bf R}_{s,p}^{\rm b}$ &received signal vector/matrix at the BS \\
		${\bf r}_{q,v}^{\rm u}$/${\bf R}_{q}^{\rm u}$ &received signal vector/matrix at the UE \\
		${\bf s}_p^{\rm u}$/${\bf s}_p^{\rm b}$& $p$-th pilot vector transmitted by the UE/BS\\
		${\rm t}_k$&direction vector of the LoS path for ${{\bf{H}}_{{{\rm ur}},k}}$ \\
		${\bf W}_{{\rm u}}$/${\bf W}_{{\rm b}}$& combiner at the UE/precoder at the BS \\
		${\bf y}_{k,s,p}^{\rm b}$/${\bf Y}_{k,s}^{\rm b}$/${\bf Y}_{k}^{\rm b}$& BS's received signal from RIS$_k$ \\
		${\bf y}_{k,q}^{\rm u}$/${\bf y}_{k}^{\rm u}$/${\bf Y}_{k}^{\rm u}$& UE's received signal from RIS$_k$ \\
		$\beta_{k,l}^{{\rm ur}}$/$\beta_{k,l}^{{\rm rb}}$ & small-scale gain of path $l$ in ${{\bf{H}}_{{{\rm ur}},k}}$ / ${{\bf{H}}_{{{\rm rb}},k}}$ \\
		$\varepsilon _0$ &penetration loss for the blocked direct channel \\
		${\theta _{k,{{0}}}^{{\rm rb},\rm{A}}}$ &AoA of the LoS link at the BS \\
		$\Theta _{k,{{0}}}^{{\rm rb},\rm{A}}$&$\Theta _{k,{{0}}}^{{\rm rb},\rm{A}} = \pi \sin \theta _{k,{{0}}}^{{\rm rb},\rm{A}}$\\
		$\theta _{k,{{0}}}^{{{\rm rb}},{\rm D}}$/$\phi _{k,{{0}}}^{{{\rm rb}},{\rm D}}$& horizontal/vertical AoD of the LoS link at the RIS \\
		$\Theta _{k,{{0}}}^{{{\rm rb}},{\rm D}} $ &$\Theta _{k,{{0}}}^{{{\rm rb}},{\rm D}} = \pi \sin \phi _{k,{{0}}}^{{{\rm rb}},{\rm D}}\sin \theta _{k,{{0}}}^{{{\rm rb}},{\rm D}}$\\
		$\Phi _{k,{{0}}}^{{{\rm rb}},{\rm D}}$&  $\Phi _{k,{{0}}}^{{{\rm rb}},{\rm D}} = \pi \cos \phi _{k,{{0}}}^{{{\rm rb}},{\rm D}}$\\
		${\bm\gamma}_k$/${\bm \Gamma}_k$&reflection vector/matrix of the RIS$_k$ \\
		${{\eta _{\rm b}}}$/${\eta _{\rm u}}$/${\eta_{{\rm h},k}}$/$ {\eta_{{\rm v},k}}$&oversampling factor of predefined grid \\
		$\kappa_{\rm ur}$/$\kappa_{\rm rb}$&Rician factor in the UE--RIS/RIS--BS channel \\
		$\lambda$&wavelength \\
		$\rho_0$/$\rho_k$& large-scale attenuation of ${\bf H}_0$/${\bf H}_k$\\
		$\tau $&stopping threshold for Algorithm 1 \\
		$  \xi_{k,l,v}  $ & cascaded path gain provided by RIS$_k$ \\
		\bottomrule
	\end{tabular}}
\end{table}



\section{System Model}\label{sec:2}
\begin{figure}
	\centering
	\includegraphics[width=0.5\textwidth]{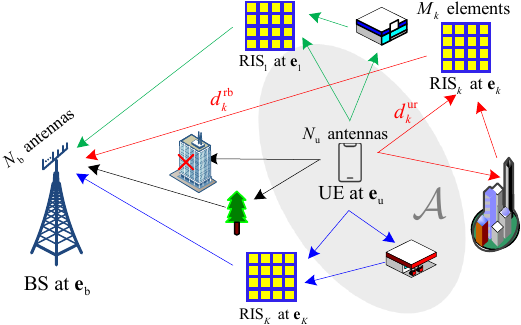}
    \caption{{Multi-RIS-assisted MIMO system, where lines of different colors represent distinct channel components through which the UE's signals reach the BS.}} 
	\label{Fig.system-model}
\end{figure}

In this section, we introduce the multi-RIS-assisted time division duplex MIMO system under consideration. As illustrated in Fig. \ref{Fig.system-model}, a BS with $N_{\rm b}$ antennas, located at ${{\bf{e}}_{\rm b}} = {\left[ {{x_{\rm b}},{y_{\rm b}},{z_{\rm b}}} \right]^T}$, provides communication services to a UE with $N_{\rm u}$ antennas. The UE is positioned at ${{\bf{e}}_{\rm u}} = {\left[ {{x_{\rm u}},{y_{\rm u}},{z_{\rm u}}} \right]^T}$. Both the BS and the UE are equipped with uniform linear arrays (ULAs). The system is assisted by $K$ RISs\footnote{{Coordination and timing synchronization among RISs are crucial for achieving the full potential of this technology, as investigated in \cite{Multi-RIS-1}. However, since these issues are beyond the scope of this work, we assume that all RISs are perfectly synchronized with the BS via solutions utilizing the global navigation satellite system \cite{GNSS}.}}, with each RIS$_k$ ($k\in {\mathcal K} \triangleq \{1,2,\ldots,K\}$) mounted at ${{\bf{e}}_k} = {\left[ {{x_k},{y_k},{z_k}} \right]^T}$. Each RIS consists of an ${M_k}$-element uniform planar array (UPA), which has ${M_{{\rm v},k}}$ elements in the vertical direction and ${M_{{\rm h},k}}$ elements in the horizontal direction. The LoS distances for the UE--RIS$_k$, RIS$_k$--BS, and UE--BS channels are $d^{\rm ur}_k$, $d^{\rm rb}_k$, and $d_0$, respectively.

\subsection{Channel Model}

The end-to-end channel ${{\bf{H}} \in {{\mathbb C}^{{N_{\rm b}} \times {N_{\rm u}}}}}$, composed of the direct UE--BS channel and the cascaded UE--RISs--BS channel, can be expressed as
\begin{equation}
	{\bf{H}} = {\rho _0}{{\bf{H}}_{{0}}} + \sum\limits_{k = 1}^K {{\rho _k}{{\bf{H}}_{{{\rm rb}},k}}{\rm diag}{\left( {{{\bm{\gamma }}_k}} \right)}{{\bf{H}}_{{{\rm ur}},k}}}  = \sum\limits_{k = 0}^K {{\rho _k}{{\bf{H}}_k}},
\end{equation}
where the second equation is derived from the notation ${{\bf{H}}_k} \triangleq {{\bf{H}}_{{{\rm rb}},k}}{\rm diag}{\left( {{{\bm{\gamma }}_k}} \right)}{{\bf{H}}_{{{\rm ur}},k}}$ for the cascaded UE--RIS$_k$--BS channel. Here, ${\rho _0} $ and ${\rho _k}$ represent the large-scale attenuation of the direct channel ${{\bf{H}}_{{0}}}$ and cascaded channel ${\bf H}_k$, respectively. In a typical scenario where the direct channel is blocked and RISs are mounted within the LoS of the BS and UE, ${\rho _0} $ and ${\rho _k} $ can be respectively given by
\begin{equation}
	{\rho _0} = \frac{{\lambda {\varepsilon _0}}}{{4\pi d_0}} ~{\rm and}~ {\rho _k} = \frac{\lambda }{{4\pi d^{{\rm ur}}_k}}\frac{\lambda }{{4\pi d^{{\rm rb}}_k}},
\end{equation}
where $\lambda$ is the wavelength, and $\varepsilon _0$ is the penetration loss for the blocked direct channel. In ${\bf H}_k$, ${{{\bm{\gamma }}_k}} = {[ {{e^{j{\gamma _{k,1}}}}, \ldots ,{e^{j{\gamma _{k,{M_k}}}}}} ]^T}\in{{\mathbb C}^{M_k \times 1}}$ is the reflection coefficient vector of RIS$_k$, ${{\bf{H}}_{{{\rm ur}},k}}\in {{\mathbb C}^{M_k\times {N_{\rm u}} }}$ and ${{\bf{H}}_{{{\rm rb}},k}}\in {{\mathbb C}^{{N_{\rm b}} \times M_k}}$ are the UE--RIS$_k$ and RIS$_k$--BS channels, respectively.

In the considered scenario, we adopt the Rayleigh fading for the direct channel ${\bf H}_0$ due to the blocked LoS, and use the Rician channel for ${\bf H}_{{\rm ur},k}$ and ${\bf H}_{{\rm rb},k}$. The RIS$_k$--BS channel is modeled as
\begin{equation}\label{Eq:H-rb}
	\begin{aligned}
		{{\bf{H}}_{{{\rm rb}},k}} = \sqrt {{M_k}{N_{\rm b}}} {\left( {\sqrt {\frac{{{\kappa _{{{\rm rb}}}}}}{{{\kappa _{{{\rm rb}}}} + 1}}} {\bf H}_{{\rm rb},k}^{\rm LoS}  } + \sqrt {\frac{1}{{{\kappa _{{{\rm rb}}}} + 1}}} {\bf{H}}_{{{\rm rb}},k}^{{\rm NLoS}} \right)},
	\end{aligned}
\end{equation}
where ${\kappa _{{\rm rb}}}$ is the Rician factor; ${\bf{H}}_{{\rm rb},k}^{\rm NLoS}$ and ${\bf H}_{{\rm rb},k}^{\rm LoS}=\beta _{k,0}^{{{\rm rb}}}{{\bf{a}}_{\rm b}}{( {\Theta _{k,{0}}^{{{\rm rb}},{\rm A}}} )} {\bf{a}}_{{\rm r},k}^H{( {\Theta _{k,{{0}}}^{{{\rm rb}},{\rm D}},\Phi _{k,{{0}}}^{{{\rm rb}},{\rm D}}} )}$ are the non-LoS (NLoS) and LoS channel components, respectively\footnote{{In this study, channel components are expressed by the product of array response vectors because the far-field with planar waves propagation is considered. In near-field cases, where spherical waves are prevalent, the channel structure is different. Extending this study to near-field scenarios necessitates a fresh analysis of the channel structure characteristics, followed by the design of appropriate RIS configurations tailored to these characteristics to customize a sparse channel.}}. In ${\bf H}_{{\rm rb},k}^{\rm LoS}$, ${\beta _{k,0}^{{\rm rb}}}$ is the normalized gain of the LoS path set as $1$, ${{{\bf{a}}_{\rm b}}}(\cdot)$ and ${\bf a}_{{\rm r},k}(\cdot)$ are the array response vectors at the BS and RIS$_k$, respectively. Considering the far-field scenario, the array response for UPA can be decomposed into that of ULA as ${\bf a}_{{\rm r},k}(\Theta,\Phi) = {\bf a}_{{\rm v},k}(\Phi)\otimes {\bf a}_{{\rm h},k}(\Theta)$. The general array response vector for a $N$-element ULA is given by ${\bf a}(X) = \frac{1}{\sqrt{N}}{\left[1,e^{jX},\ldots,e^{j(N-1)X}\right]}^T$. The array response vector incorporates $\Theta _{k,{{0}}}^{{\rm rb},\rm{A}} = \pi \sin \theta _{k,{{0}}}^{{\rm rb},\rm{A}}$ with ${\theta _{k,{{0}}}^{{\rm rb},\rm{A}}}$ being the AoA of the LoS link at the BS, and $\Theta _{k,{{0}}}^{{{\rm rb}},{\rm D}} = \pi \sin \phi _{k,{{0}}}^{{{\rm rb}},{\rm D}}\sin \theta _{k,{{0}}}^{{{\rm rb}},{\rm D}}$ and $\Phi _{k,{{0}}}^{{{\rm rb}},{\rm D}} = \pi \cos \phi _{k,{{0}}}^{{{\rm rb}},{\rm D}}$ with $\theta _{k,{{0}}}^{{{\rm rb}},{\rm D}}$ and $\phi _{k,{{0}}}^{{{\rm rb}},{\rm D}}$ being the horizontal and vertical AoDs of the LoS link at the RIS, respectively. The AoA and AoDs of the LoS link in the RIS$_k$--BS channel are known at the BS.

The geometry multipath channel model \cite{AAM} is adopted for the NLoS channel component, expressed as
\begin{equation}\label{Eq:H-rb-NLoS}
	{\bf{H}}_{{{\rm rb}},k}^{{\rm NLoS}} = \sum\limits_{l = 1}^{{L_{k}^{\rm rb}}} { \frac{\beta _{k,l}^{{{\rm rb}}}}{{\sqrt {{L_{k}^{\rm rb}}} }} {{\bf{a}}_{\rm b}}{\left( {\Theta _{k,l}^{{{\rm rb}},{\rm A}}} \right)} {\bf{a}}_{{\rm r},k}^H{\left( {\Theta _{k,l}^{{{\rm rb}},{\rm D}},\Phi _{k,l}^{{{\rm rb}},{\rm D}}} \right)}} ,
\end{equation}
where ${L^{\rm rb}_k}$ is the number of NLoS paths and $\beta _{k,l}^{{\rm rb}}\sim {\mathcal {CN}}(0,1)$ is the small-scale path gain of the $l$-th path. The definitions for ${\Theta _{k,l}^{{\rm rb},\rm{A}}}$, $\Theta _{k,l}^{{{\rm rb}},{\rm D}}$, and $\Phi _{k,l}^{{{\rm rb}},{\rm D}}$ are similar to those of ${\Theta _{k,0}^{{{\rm rb}},{\rm A}}}$, $\Theta _{k,0}^{{{\rm rb}},{\rm D}}$, and $\Phi _{k,0}^{{{\rm rb}},{\rm D}}$, respectively. Substituting \eqref{Eq:H-rb-NLoS} into \eqref{Eq:H-rb}, the RIS$_k$--BS channel is unified as
\begin{equation}\label{Eq:H-rb-1}
	{{\bf{H}}_{{{\rm rb}},k}} = \sum\nolimits_{l = 0}^{{L_{k}^{\rm rb}}} {g_{k,l}^{{{\rm rb}}}{{\bf{a}}_{\rm b}}{( {\Theta _{k,l}^{{{\rm rb}},{\rm A}}} )}{\bf{a}}_{{\rm r},k}^H{( {\Theta _{k,l}^{{{\rm rb}},{\rm D}},\Phi _{k,l}^{{{\rm rb}},{\rm D}}} )}} ,
\end{equation}
where
\begin{equation}
	g_{k,0}^{{{\rm rb}}} = \sqrt {\frac{{{M_k}{N_{\rm b}}{\kappa _{{{\rm rb}}}}}}{{{\kappa _{{{\rm rb}}}} + 1}}} \beta _{k,0}^{{{\rm rb}}}\ {\rm and}\ g_{k,l}^{{{\rm rb}}} = \sqrt {\frac{{{M_k}{N_{\rm b}}}}{{\left( {{\kappa _{{{\rm rb}}}} + 1} \right){L_{k}^{\rm rb}}}}} \beta _{k,l}^{{{\rm rb}}}.
\end{equation}
Following \eqref{Eq:H-rb-1}, the UE--RIS$_k$ channel is expressed as
\begin{equation}
	{{\bf{H}}_{{{\rm ur}},k}} = \sum\nolimits_{l = 0}^{{L_{k}^{\rm ur}}} {g_{k,l}^{{{\rm ur}}}{{\bf{a}}_{{\rm r},k}}{\left( {\Theta _{k,l}^{{{\rm ur}},{\rm A}},\Phi _{k,l}^{{{\rm ur}},{\rm A}}} \right)}{\bf{a}}_{\rm u}^H{\left( {\Theta _{k,l}^{{{\rm ur}},{\rm D}}} \right)}} ,
\end{equation}
where ${\bf{a}}_{\rm u}(\cdot)$ is the array response vector at the UE, ${{L_{k}^{\rm ur}}}$ is the number of NLoS paths, and the remaining parameters including $g_{k,l}^{{\rm ur}}$, $\kappa_{{{\rm ur}}}$, ${\Theta _{k,l}^{{\rm ur},\rm{A}}}$, ${\Phi _{k,l}^{{\rm ur},\rm{A}}}$, and ${\Theta _{k,l}^{{\rm ur},\rm{D}}}$ are defined in a manner similar to that in the RIS$_k$--BS channel.

\subsection{SE}
In downlink transmission, the received signal is given by
\begin{equation}
	{\bf{y}} = {\bf{W}}_{\rm u}^H{{\bf{H}}^H}{{\bf{W}}_{\rm b}}{\bf{s}} + {\bf{W}}_{\rm u}^H{\bf{n}},
\end{equation}
where ${\bf{s}} = {\left[ {{s_1},{s_2}, \ldots ,{s_{{N_{\rm u}}}}} \right]^T} $ with ${\mathbb E}\left\{ {{\bf{s}}{{\bf{s}}^H}} \right\} = {\bf{I}}$ represents the transmit signal, ${{\bf{W}}_{\rm b}} \in {{\mathbb C}^{{N_{\rm b}} \times {N_{\rm u}}}}$ is the precoder at the BS, ${{\bf{W}}_{\rm u}} \in {{\mathbb C}^{{N_{\rm u}} \times {N_{\rm u}}}}$ is the combiner at the UE, and ${\bf{n}} \in {{\mathbb C}^{{N_{\rm u}}\times 1}}\sim {\mathcal {CN}}(0,\sigma^2{\bf I})$ is the additive white Gaussian noise with a noise power of $\sigma^2$. The power constraint at the BS is $\| {{{\bf{W}}_{\rm b}}{\bf{s}}} \|_F^2 \le {P_{\rm b}}$, where ${P_{\rm b}}$ denotes the transmit power.

The downlink SE can be expressed as
\begin{equation}\label{Eq:R}
	 {\log _2}\det{( {{\bf{I}} + {\bf{W}}_{\rm u}^H{{\bf{H}}^H}{{\bf{W}}_{\rm b}}{\bf{W}}_{\rm b}^H{\bf{H}}{{\bf{W}}_{\rm u}}/{\sigma ^2}} )}.
\end{equation}
Maximizing downlink SE involves a joint design problem for the precoder, reflection vector, and combiner. In conventional systems without RIS, optimal precoder and combiner designs can be achieved through the singular value decomposition (SVD) of ${\bf H}^H$ \cite{MIMO_OFDM}. Given the SVD of ${\bf H}^H$ as $ {{\bf{V}}_{\rm u}}{\bf{\Lambda V}}_{\rm b}^H$, the optimal combiner is ${{\bf{W}}_{\rm u}} = {{\bf{V}}_{\rm u}}$ and the optimal precoder is ${{\bf{W}}_{\rm b}} = {{\bf{V}}_{\rm b}}{{\bf{P}}^{1/2}}$, where the diagonal matrix ${\bf{P}}$ is the water-filling power allocation depending on the singular value matrix ${\bf \Lambda}$ and the transmit power $P_{\rm b}$. Applying the SVD transceiver, \eqref{Eq:R} can be rewritten as
\begin{equation}\label{Eq:R-1}
	\sum\nolimits_{i = 1}^Q {{\log }_2}{\left( {1 + {{{p_i}{\lambda _i}}}/{{{\sigma ^2}}}} \right)} ,
\end{equation}
where $Q$ is the channel rank, $p_i$ is the $i$-th diagonal element of ${\bf P}$, and $\{{\lambda _i}\}^Q_{i=1}$are the singular values that should be optimized by adjusting the reflection coefficients of RISs based on instantaneous CSI. The SE maximization problem requires only the end-to-end channel $\bf H$ in conventional systems, but necessitates multiple channel components including the direct channel ${\bf H}_0$ and segmented channels $\{{\bf H}_{{\rm rb},k},{\bf H}_{{\rm ur},k}\}$ in multi-RIS-assisted systems. Given that the segmented channels are high-dimensional and low-cost RIS typically lacks channel estimation capabilities, obtaining instantaneous CSI for RIS reflection design is extremely challenging. Moreover, optimizing RISs to maximize SE in \eqref{Eq:R-1} is prohibitively complicated because singular values cannot be expressed in closed-forms and must be derived from channel decomposition whenever the reflection coefficients change.

Achieving optimal SE requires perfect CSI and the most complex optimization algorithm. Considering the resources required for channel estimation and RIS optimization, achieving this goal in practical wireless systems is impractical. We propose a sub-optimal solution with low-complexity channel estimation and RIS optimization by customizing the channel.


\section{Channel Separation and Extraction}\label{sec:3}
Unlike single-RIS-assisted scenarios, channel estimation in multi-RIS-assisted systems presents additional challenges, as reflection channels cascaded by different RISs become intertwined during the channel training stage. This section introduces a training strategy for RIS configurations aimed at isolating different channel components. Additionally, we briefly discuss a parameter extraction algorithm based on the Newtonized Orthogonal Matching Pursuit (NOMP), which is utilized to estimate the decoupled channel components.

\subsection{Channel Separation}\label{sec:3-1}

\begin{figure}
	\centering
	\includegraphics[width=0.5\textwidth]{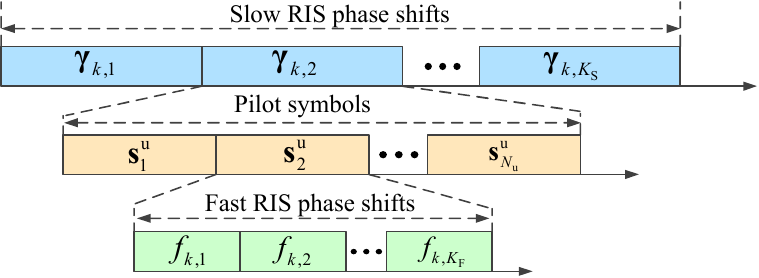}
	\caption{Frame structure for the uplink channel estimation.}
	\label{Fig.frame}
\end{figure}

During the uplink channel estimation process, RISs are required to adjust their reflection coefficients in conjunction with the transmission of uplink pilots. The slow- and fast-varying RIS phase shifts \cite{slow-fast} are utilized to design the uplink training protocol. \textcolor{black}{While the slow-varying RIS phase shifts are employed for regular beam training, we draw inspiration from the approach in \cite{slow-fast}, which uses temporal phase shifts to eliminate interpath interference, and design fast-varying RIS phase shifts to isolate channel components.} As depicted in Fig. \ref{Fig.frame}, RIS$_k$ is configured with $K_{\rm S}$ slow reflection vectors $\{ {{{\bm{\gamma }}_{k,s}}} \}_{s = 1}^{{K_{\rm S}}}$. \textcolor{black}{All RISs perform a synchronized switch of their slow-varying reflection vectors, resulting in a time overhead identical to that of single RIS-assisted systems.} The UE periodically transmits $N_{\rm u}$ pilot symbols $\{ {{{\bf{s}}^{\rm u}_{p}}} \}_{p = 1}^{{N_{\rm u}}}$ for each slow RIS reflection vector. Beyond this regular training stage, we propose successively adding $K_{\rm F}$ fast common reflections $\{ {{f_{k,v}}} \}_{v = 1}^{{K_{\rm F}}}$ to each RIS$_k$ during each pilot symbol. Since the common reflection applies to all elements, the reflection matrix of RIS$_k$ can be decoupled as $ {f_{k,v}} \times {\rm diag}( {{{\bm{\gamma }}_{k,s}}} )$. For simplification, we use the DFT matrix, a special case in  \cite{slow-fast}, to define a fast reflection matrix ${\bf{F}} = [ {{{\bf{f}}_0},{{\bf{f}}_1}, \ldots ,{{\bf{f}}_K}} ] \in {{\mathbb C}^{{K_{\rm F}} \times {K_{\rm F}}}}$, where ${{K_{\rm F}} = K + 1}$, ${{\bf{f}}_0} = {{\bf{1}}_{{K_{\rm F}}}}$, and ${{\bf{f}}_k} = {\left[ {{f_{k,1}}, \ldots ,{f_{k,{K_{\rm F}}}}} \right]^T}$. The channel for the $v$-th fast reflection in the $s$-th slow reflection is denoted by ${\bf{H}}( {s,v} )$, where $v = 1, \ldots ,{K_{\rm F}}$ and $s = 1, \ldots ,{K_{\rm S}}$.

For the $v$-th fast reflection in the $p$-th pilot of the $s$-th slow reflection, the received signal at the BS is given by
\begin{equation}\label{Eq:r-1}
	\begin{aligned}
		&{\bf{r}}_{s,p,v}^{\rm b} = {\bf{H}}\left( {s,v} \right){{\bf{s}}^{\rm u}_{p}} + {{\bf{n}}_{s,p,v}}\\
		&= {\left( {{\rho _0}{{\bf{H}}_0} + \sum\limits_{k = 1}^K {{f_{k,v}}{\rho _k}{{\bf{H}}_{{{\rm rb}},k}}{\rm diag}{\left( {{{\bm{\gamma }}_{k,s}}} \right)}{{\bf{H}}_{{{\rm ur}},k}}} } \right)}{{\bf{s}}^{\rm u}_{p}} + {{\bf{n}}_{s,p,v}}.
	\end{aligned}
\end{equation}
By denoting ${{{\bf{H}}_{k,s}}}={{{\bf{H}}_{{{\rm rb}},k}}{\rm diag}{\left( {{{\bm{\gamma }}_{k,s}}} \right)}{{\bf{H}}_{{{\rm ur}},k}}}$ for $k\in {\mathcal K}$ and ${{{\bf{H}}_{0,s}}}={{f_{0,v}}{{\bf{H}}_{0}}}$, stacked signals for $K_{\rm F}$ fast reflections ${\bf{R}}_{s,p}^{\rm b} \triangleq [ {{\bf{r}}_{s,p,1}^{\rm b}, \ldots ,{\bf{r}}_{s,p,{K_{\rm F}}}^{\rm b}} ]$ can be rewritten as
\begin{equation}
		{\bf{R}}_{s,p}^{\rm b} =\left[ {{\rho _0}{{\bf{H}}_{0,s}}{{\bf{s}}^{\rm u}_{p}}, \ldots ,{\rho _K}{{\bf{H}}_{K,s}}{{\bf{s}}^{\rm u}_{p}}} \right]{{\bf{F}}^T} + {{\bf{N}}_{s,p}}.
\end{equation}
Multiplying ${\bf{R}}_{s,p}^{\rm b}$ with ${\bf F}^*$, we obtain
\begin{equation}
	\begin{aligned}
		{\bf{R}}_{s,p}^{\rm b}{{\bf{F}}^*}  &= \left[ {{\rho _0}{{\bf{H}}_{0,s}}{{\bf{s}}^{\rm u}_{p}}, \ldots ,{\rho _K}{{\bf{H}}_{K,s}}{{\bf{s}}^{\rm u}_{p}}} \right]{{\bf{F}}^T}{{\bf{F}}^*} + {{\bf{N}}_{s,p}}{{\bf{F}}^*}\\
	&	 \triangleq  \left[ {{\bf{y}}_{0,s,p}^{\rm b},{\bf{y}}_{1,s,p}^{\rm b}, \ldots ,{\bf{y}}_{K,s,p}^{\rm b}} \right].
	\end{aligned}
\end{equation}
Since ${{\bf{F}}^T}{{\bf{F}}^*} = {K_{\rm F}}{{\bf{I}}_{{K_{\rm F}} \times {K_{\rm F}}}}$, signals from different channel components can be separated. For instance, the signal from the direct UE--BS channel is given by
\begin{equation}
	{\bf{y}}_{0,s,p}^{\rm b} = {K_{\rm F}}{\rho _0}{{\bf{H}}_{0,s}}{{\bf{s}}^{\rm u}_{p}} + {{\bf{N}}_{s,p}}{\bf{f}}_0^*,
\end{equation}
and the signal component from the reflection channel cascaded by RIS$_k$ is given by
\begin{equation}\label{Eq:yb-k-s-p}
	{\bf{y}}_{k,s,p}^{\rm b} = {K_{\rm F}}{\rho _k}{{\bf{H}}_{{{\rm rb}},k}}{\rm diag}\left( {{{\bm{\gamma }}_{k,s}}} \right){{\bf{H}}_{{{\rm ur}},k}}{{\bf{s}}^{\rm u}_{p}} + {{\bf{N}}_{s,p}}{\bf{f}}_k^*.
\end{equation}

\subsection{Complete Channel Extraction}\label{sec:3-2}

Building on the proposed fast reflection training, the uplink channel estimation problem in systems assisted by $K$ RISs can be divided into $K+1$ independent sub-problems. As channel estimation in direct MIMO links is well-explored, we focus on cascaded links. Various parameter extraction algorithms can be applied to estimate each UE--RIS--BS channel. We briefly introduce an algorithm based on NOMP to illustrate the complexity of complete channel estimation.\footnote{The NOMP algorithm has been applied to estimate cascaded channels in mmWave systems \cite{NOMP-RIS}, assuming the BS–RIS link as a deterministic LoS channel. The parameter extraction process in \cite{NOMP-RIS} resembles that in conventional systems without RIS, as the BS–RIS channel with only the LoS path is considered known. In general cases, where the BS–RIS link includes multiple NLoS paths, parameter extraction becomes more complex.}

Take the UE--RIS$_k$--BS channel as an example. Stacking the signal component from \eqref{Eq:yb-k-s-p} for ${N_{\rm u}}$ pilots yields
\begin{equation}
	\begin{aligned}
		&{\bf{Y}}_{k,s}^{\rm b} = {\left[ {{\bf{y}}_{k,s,1}^{\rm b}, \ldots ,{\bf{y}}_{k,s,{N_{\rm u}}}^{\rm b}} \right]}\\
		 &= {K_{\rm F}}{\rho _k}{{\bf{H}}_{{{\rm rb}},k}}{\rm diag}{\left( {{{\bm{\gamma }}_{k,s}}} \right)}{{\bf{H}}_{{{\rm ur}},k}}{{\bf{S}}_{\rm u}}+{{\bf{N}}_s}{\left( {{{\bf{I}}_{{N_{\rm u}} \times {N_{\rm u}}}} \otimes {\bf{f}}_k^*} \right) },
	\end{aligned}
\end{equation}
where ${{\bf{Y}}_{k,s}^{\rm b} \in {{\mathbb C}^{{N_{\rm b}} \times {N_{\rm u}}}}}$, ${{\bf{N}}_s} = \left[ {{{\bf{N}}_{s,1}}, \ldots ,{{\bf{N}}_{s,{N_{\rm u}}}}} \right] \in {{\mathbb C}^{{N_{\rm b}} \times {K_{\rm F}}{N_{\rm u}}}}$, and ${\bf S}_{\rm u}=[{\bf s}^{\rm u}_{1},\ldots,{\bf s}^{\rm u}_{N_{\rm u}}]$ is the UE's pilot matrix satisfying ${\bf S}_{\rm u}{\bf S}_{\rm u}^H = P_{\rm u}{\bf I}$. Signals for $K_{\rm S}$ slow RIS reflections are then aggregated as
\begin{equation}\label{Eq:Yb-m}
	{\bf{Y}}_k^{\rm b} = \left[ {{\rm vec}\left( {{\bf{Y}}_{k,1}^{\rm b}} \right), \ldots ,{\rm vec}\left( {{\bf{Y}}_{k,{K_{\rm S}}}^{\rm b}} \right)} \right] \in {{\mathbb C}^{{N_{\rm b}}{N_{\rm u}} \times {K_{\rm S}}}}.
\end{equation}
Applying the vectorization identity\footnote{ ${\rm vec}( {{\bf{A}}{\rm diag}\left( {\bf{x}} \right){\bf{B}}} ) = ( {{{\bf{B}}^T}  \oplus  {\bf{A}}} ){\bf{x}}$} to \eqref{Eq:Yb-m} and denoting ${{\bm{\Gamma }}_k}=[ {{{\bm{\gamma }}_{k,1}}, \ldots ,{{\bm{\gamma }}_{k,{K_{\rm S}}}}} ]$, we can rewrite it as
\begin{multline}
	{\bf{Y}}_k^{\rm b}={K_{\rm F}}{\rho _k}{\left( {\left( {{\bf{S}}_{\rm u}^T{\bf{H}}_{{{\rm ur}},k}^T} \right) \oplus {{\bf{H}}_{{{\rm rb}},k}}} \right)}{{\bf{\Gamma }}_k} \\
+ {\rm vec}{\left( {{{\bf{N}}_s}{\left( {{{\bf{I}}_{{N_{\rm u}} \times {N_{\rm u}}}} \otimes {\bf{f}}_k^*} \right)}} \right)}.
\end{multline}
Considering that the RIS$_k$--BS channel and the UE--RIS$_k$ channel each comprise ${{L_{k}^{\rm rb}} + 1}$ and ${{L_{k}^{\rm ur}} + 1}$ paths respectively, there are ${( {{L_{k}^{\rm ur}} + 1} )}{( {{L_{k}^{\rm rb}} + 1} )}$ distinct paths cascaded by RIS$_k$. Estimating these paths independently increases the iterations of the parameter extraction algorithm and introduces angular ambiguity in the AoA and AoD of the RIS \cite{YXLin}. Thus, we propose estimating individual paths of the RIS$_k$--BS channel and the UE--RIS$_k$ channel instead.

\begin{figure*}
	\centering
	\includegraphics[width=1\textwidth]{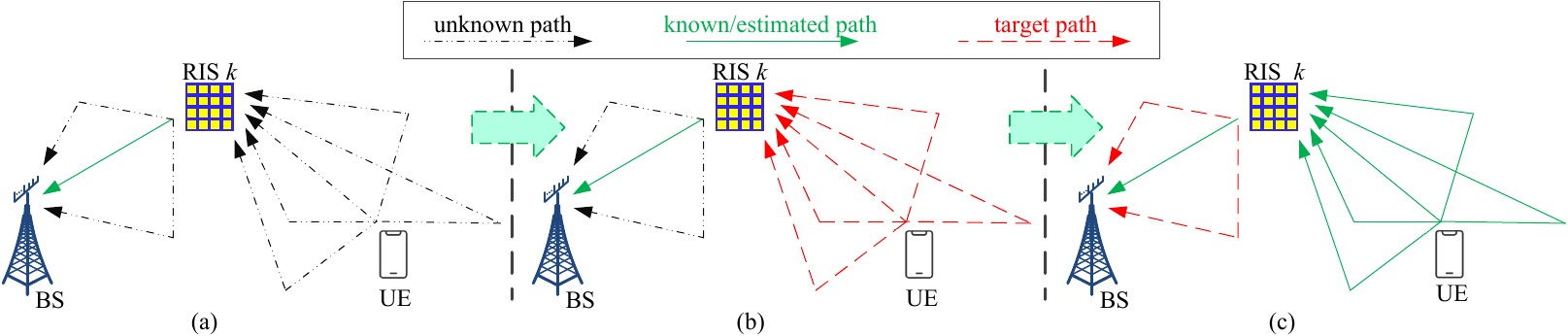}

    \caption{Overview of the parameter extraction process for the UE--RIS$_k$--BS channel: (a) Determination of the LoS path in the RIS$_k$--BS channel, made predictable by the stationary positions of the BS and RIS$_k$; (b) Priority estimation of cascaded paths incorporating the LoS path within the RIS$_k$--BS channel; (c) Subsequent estimation of the remaining NLoS paths in the RIS$_k$--BS channel.}

	\label{Fig.NOMP}
\end{figure*}

In typical scenarios where the locations of the BS and RIS are fixed, the directional parameters of the LoS path in the RIS$_k$--BS channel are deterministic, as illustrated in Fig. \ref{Fig.NOMP}(a). The normalized channel component of path $i$ in the UE--RIS$_k$ channel and path $j$ in the RIS$_k$--BS channel are denoted as ${{\bf{U}}_{k,i}} = {{\bf{a}}_{{\rm r},k}}( {\Theta _{k,i}^{{{\rm ur}},{\rm A}},\Phi _{k,i}^{{{\rm ur}},{\rm A}}} ){\bf{a}}_{\rm u}^H( {\Theta _{k,i}^{{{\rm ur}},{\rm D}}} )$ and ${{\bf{B}}_{k,j}} = {{\bf{a}}_{\rm b}}( {\Theta _{k,j}^{{{\rm rb}},{\rm A}}} ){\bf{a}}_{{\rm r},k}^H( {\Theta _{k,j}^{{{\rm rb}},{\rm D}},\Phi _{k,0}^{{{\rm rb,D}}}} )$, respectively. Given that ${{\bf{B}}_{k,0}}$ is deterministic, prioritizing the extraction of paths in the UE--RIS$_k$ channel is key for reconstructing the cascaded paths that include the LoS path in the RIS$_k$--BS channel, as shown in Fig. \ref{Fig.NOMP}(b). Assuming parameters of paths $\{0,1\ldots,i-1\}$ in the UE--RIS$_k$ channel are already extracted, the maximum likelihood (ML) estimation of parameters for path $i$ is obtained by \eqref{Eq:NOMP-UE},
\begin{figure*}
	\begin{equation}\label{Eq:NOMP-UE}
		\left\{ {\hat g_{k,i}^{{{\rm ur}}},\hat \Theta _{k,i}^{{{\rm ur}},{\rm A}},\hat \Phi _{k,i}^{{{\rm ur}},{\rm A}},\hat \Theta _{k,i}^{{{\rm ur}},{\rm D}}} \right\} = \mathop {\arg \min }\limits_{\left\{ {g_{k,i}^{{{\rm ur}}},\Theta _{k,i}^{{{\rm ur}},{\rm A}},\Phi _{k,i}^{{{\rm ur}},{\rm A}},\Theta _{k,i}^{{{\rm ur}},{\rm D}}} \right\}} \left\| {{{\bf{Z}}_{0,i - 1}} - {K_{\rm F}}{\rho _k}\left( {g_{k,i}^{{{\rm ur}}}\left( {{\bf{S}}_{\rm u}^T{\bf{U}}_{k,i}^T} \right) \oplus g_{k,0}^{{{\rm rb}}}{{\bf{B}}_{k,0}}} \right){{\bf{\Gamma }}_k}} \right\|_F^2.
	\end{equation}
\hrulefill
\end{figure*}
where
\begin{equation*}
	{{\bf{Z}}_{0,i - 1}} = {\bf{Y}}_k^{\rm b} - {K_{\rm F}}{\rho _k} {\left( { {\Big( {\sum_{i' = 0}^{i - 1} {\hat g_{k,i'}^{{{\rm ur}}}{\bf{S}}_{\rm u}^T{\bf{\hat U}}_{k,i'}^T} } \Big)} \oplus g_{k,0}^{{{\rm rb}}}{{\bf{B}}_{k,0}}} \right)}{{\bf{\Gamma }}_k}
\end{equation*}
represents the residual power and ${\sum_{i' = 0}^{i - 1} {\hat g_{k,i'}^{{{\rm ur}}}{\bf{S}}_{\rm u}^T{\bf{\hat U}}_{k,i'}^T} }$ denotes the estimated components of paths $\{0,1,\ldots,i-1\}$. This problem can be solved using the NOMP algorithm, the steps of which (including exhaustive coarse estimation, single refinement, cyclic refinement, and gains update) are detailed in \cite{NOMP-Or} and are not reiterated here.

Once parameters for ${L_{k}^{\rm ur}}+1$ paths are obtained, the estimated UE--RIS$_k$ channel is given by ${{\hat{\bf H}}_{{{\rm ur}},k}} = \sum\nolimits_{i = 0}^{{L_{k}^{\rm ur}}} {\hat g_{k,i}^{{{\rm ur}}}{{{\hat{\bf U}}}_{k,i}}} $. We then proceed to estimate the RIS$_k$--BS channel, as depicted in Fig. \ref{Fig.NOMP}(c). The ML estimation of parameters for path $j$ in the BS–RIS$_k$ channel is formulated as \eqref{Eq:NOMP-BS},
\begin{figure*}
	\begin{equation}\label{Eq:NOMP-BS}
		\left\{ {\hat g_{k,j}^{{{\rm rb}}},\hat \Theta _{k,j}^{{{\rm rb}},{\rm D}},\hat \Phi _{k,j}^{{{\rm rb}},{\rm D}},\hat \Theta _{k,j}^{{{\rm rb}},{\rm A}}} \right\} = \mathop {\arg \min }\limits_{\left\{ {g_{k,j}^{{{\rm rb}}},\Theta _{k,j}^{{{\rm rb}},{\rm D}},\Phi _{k,j}^{{{\rm rb}},{\rm D}},\Theta _{k,j}^{{{\rm rb}},{\rm A}}} \right\}} \left\| {{{\bf{Z}}_{j - 1,{L_{k}^{\rm ur}}}} - {K_{\rm F}}{\rho _k}\left( {\left( {{\bf{S}}_{\rm u}^T{\hat{ \bf H}}_{{{\rm ur}},k}^T} \right) \oplus g_{k,j}^{{{\rm rb}}}{{\bf{B}}_{k,j}}} \right){{\bf{\Gamma }}_k}} \right\|_F^2.
	\end{equation}
\hrulefill
\end{figure*}
where
\begin{multline*}
		{{\bf{Z}}_{j - 1,{L_{k}^{\rm ur}}}}
		 = {{\bf{Y}}_{k}^{\rm b}} \\- {K_{\rm F}}{\rho _k} {\left(  {\Big( {{\bf{S}}_{\rm u}^T{\hat{\bf H}}_{{{\rm ur}},k}^T} \Big)} \oplus {\Big( \sum_{j' = 0}^{j - 1} {\hat g_{k,j'}^{{{\rm rb}}}{{{\hat{\bf B}}}_{k,j'}}} \Big)}  \right)}{{\bf{\Gamma }}_k}.
\end{multline*}
Problem \eqref{Eq:NOMP-BS} can be approached in a similar manner to \eqref{Eq:NOMP-UE}.

The complexity of channel estimation using the NOMP method is primarily driven by the exhaustive coarse estimation of directional parameters (AoAs and AoDs) over predefined grids \cite{NOMP-Chen}. In scenarios where the number of NLoS paths $\{{L_{k}^{\rm rb}},{L_{k}^{\rm ur}}\}_{k=1}^{K}$ is known, the search complexity of this exhaustive coarse estimation can be expressed as
\begin{equation}\label{Eq:C-full}
	C = \sum\nolimits_{k = 1}^K {{\eta _{{\rm{v}},k}}{\eta _{{\rm{h}},k}}{M_k}{\left( {\left( {L_k^{{\rm{ur}}} + 1} \right){\eta _{\rm{u}}}{N_{\rm{u}}} + L_k^{{\rm{rb}}}{\eta _{\rm{b}}}{N_{\rm{b}}}} \right)}} ,
\end{equation}
where ${{\eta _{\rm b}}}$, ${\eta _{\rm u}}$, ${\eta_{{\rm h},k}}$, and $ {\eta_{{\rm v},k}}$,  are oversampling factors of predefined grids for estimating $\{{\Theta _{k,l}^{{{\rm rb}},{\rm A}}}\}_{l=1}^{L_{k}^{\rm rb}}$, $\{{\Theta _{k,l}^{{{\rm ur}},{\rm D}}} \}_{l=0}^{L_{k}^{\rm ur}}$, $\{\{{\Theta _{k,l}^{{{\rm rb}},{\rm D}}}\}_{l=1}^{L_{k}^{\rm rb}}, \{{\Theta _{k,l}^{{{\rm ur}},{\rm A}}} \}_{l=0}^{L_{k}^{\rm ur}}\}$, and $\{\{{\Phi _{k,l}^{{{\rm rb}},{\rm D}}}\}_{l=1}^{L_{k}^{\rm rb}}, \{{\Phi _{k,l}^{{{\rm ur}},{\rm A}}} \}_{l=0}^{L_{k}^{\rm ur}}\}$, respectively. 

To illustrate the search complexity, consider a special case where ${M_k} = M$, ${\eta_{{\rm v},k}} = {\eta_{{\rm h},k}} = {\eta _{\rm u}} = {\eta _{\rm b}} = \eta $, ${L_{k}^{\rm ur}} + 1 \approx {L^{{{\rm ur}}}}$, and ${L_{k}^{\rm rb}} = {L^{{{\rm rb}}}}$. In this scenario, the complexity simplifies to
\begin{equation}\label{Eq:C-sim}
	C=KM{\eta ^3}\left( {{L^{{{\rm ur}}}}{N_{\rm u}}+{L^{{{\rm rb}}}}{N_{\rm b}}} \right).
\end{equation}
\textcolor{black}{The search complexity of the NOMP method is thus proportional to the total number of RIS elements and the number of paths.} With large-scale RIS deployment, this complexity escalates due to the high-dimensional matrix computations required for ${\bf U}_{k,i}\in {\mathbb C}^{{M}\times{N_{\rm u}}}$, ${\bf B}_{k,j}\in {\mathbb C}^{{N_{\rm b}}\times M}$, and ${\bf \Gamma}_k \in {\mathbb C}^{M\times K_{\rm S}}$ at every step of the exhaustive coarse search. The overall computational load may become unsustainable, even for the BS, especially in the rich-scattering environments introduced by multiple RISs. To address this challenge, the following section discusses how RIS can be leveraged to transform the dense environment into a sparse one, thereby reducing the estimation complexity.

\section{Low-complexity CSI Acquisition}\label{sec:4}
Building on the previous section's efforts to simplify channel estimation in multi-RIS-assisted systems by separating channel components, we still face challenges in parameter extraction for all paths. This section shows how, by adjusting RIS phases with partial CSI, we can shape the end-to-end channel\footnote{We primarily focus on the reflection channel mediated by RISs, as the direct channel can be independently estimated and separated.} to be predominantly influenced by a select few paths. While this approach of tailoring the channel through RIS configuration might not yield the highest SE, it offers a more straightforward system implementation. We then introduce a low-complexity method for uplink channel estimation, aimed at obtaining the partial CSI needed for channel customization. Utilizing this tailored channel, we demonstrate how the process for downlink channel estimation can be markedly simplified.

\subsection{Channel Customization}\label{sec:4-1}

In this subsection, we customize the channel to reduce the complexity of uplink channel estimation and the overhead of downlink channel estimation. To this end, we first reformulate the reflection channel as follows:
\begin{equation}\label{Eq:H-ori}
	{\bf{H}} = \sum\nolimits_{k = 1}^K {{\rho _k}{{\bf{H}}_{{{\rm rb}},k}}{\rm diag}\left( {{{\bm{\gamma }}_k}} \right){{\bf{H}}_{{{\rm ur}},k}}}  = {{\bf{A}}_{\rm b}}{\bf{\Xi A}}_{\rm u}^H,
\end{equation}
where ${\bf{\Xi }} = {\rm diag}( {{{\bf{\Xi }}_1}, \ldots ,{{\bf{\Xi }}_K}} )$ with elements representing the cascaded path gains. Here, ${{\bf{A}}_{\rm b}} = [ {{{\bf{A}}_{{\rm b},1}}, \ldots ,{{\bf{A}}_{{\rm b},K}}} ]$ and ${{\bf{A}}_{\rm u}} = [ {{{\bf{A}}_{{\rm u},1}}, \ldots ,{{\bf{A}}_{{\rm u},K}}} ]$ are the array response matrices at the BS and the UE, respectively. In ${\bf A}_{\rm b}$ and ${\bf A}_{\rm u}$, the array response matrices corresponding to RIS$_k$ are given by
\begin{equation}
	\left\{ \begin{aligned}
		&{{\bf{A}}_{{\rm{b}},k}} = \left[ {{{\bf{a}}_{\rm{b}}}\left( {\Theta _{k,0}^{{\rm{rb}},{\rm{A}}}} \right), \ldots ,{{\bf{a}}_{\rm{b}}}\left( {\Theta _{k,L_k^{{\rm{rb}}}}^{{\rm{rb}},{\rm{A}}}} \right)} \right],\\
	&{{\bf{A}}_{{\rm{u}},k}} = \left[ {{{\bf{a}}_{\rm{u}}}\left( {\Theta _{k,0}^{{\rm{ur}},{\rm{D}}}} \right), \ldots ,{{\bf{a}}_{\rm{b}}}\left( {\Theta _{k,L_k^{{\rm{ur}}}}^{{\rm{ur}},{\rm{D}}}} \right)} \right].
\end{aligned} \right.
\end{equation}
The element in the ${(l+1)}$-th row and ${(v+1)}$-th column of ${{{\bf{\Xi }}_k}}$, for $k\in\{1,2,\ldots,K\}$, is given by
\begin{equation}
	\begin{aligned}
		{\xi _{k,l ,v }} = &{\rho _k}g_{k,l }^{{{\rm rb}}}g_{k,v }^{{{\rm ur}}}{\bf{a}}_{{\rm r},k}^H{\left( {\Theta _{k,l }^{{{\rm rb}},{\rm D}},\Phi _{k,l }^{{{\rm rb}},{\rm D}}} \right)}\\
		&\times {\rm diag}{\left( {{{\bm{\gamma }}_k}} \right)}{{\bf{a}}_{{\rm r},k}}{\left( {\Theta _{k,v }^{{{\rm ur}},{\rm A}},\Phi _{k,v}^{{{\rm ur}},{\rm A}}} \right)},
	\end{aligned}
\end{equation}
which indicates the gain of the end-to-end path cascaded by the path $l$ in the RIS$_k$--BS channel and the path $v$ in the UE--RIS$_k$ channel.

Because ${\xi _{k,l,v}}$ depends on the reflection vector of the RIS, $\bf \Xi$ is the customizable channel component. By tuning $\bf \Xi$ with dedicated $\{{\bm \gamma}_k\}_{k=1}^{K}$, the reflection channel $\bf H$ can be arbitrarily customized to meet communication requirements. In the considered scenario, where LoS paths exist in both the RIS$_k$--BS and UE--RIS$_k$ channels, we design the reflection vector to enhance these LoS paths. Assuming that the AoAs and AoDs of the LoS path can be obtained, the reflection vector maximizing the cascaded LoS path is given by
\begin{equation}\label{Eq:RIS-design}
	\textcolor{black}{{{\bm{\gamma }}_k} = {M_k}{{\bf{a}}_{{\rm r},k}}{\left( {\Theta _{k,0}^{{{\rm rb}},{\rm D}},\Phi _{k,0}^{{{\rm rb}},{\rm D}}} \right)} \odot {\bf{a}}_{{\rm r},k}^*{\left( {\Theta _{k,0}^{{{\rm ur}},{\rm A}},\Phi _{k,0}^{{{\rm ur}},{\rm A}}} \right)}}.
\end{equation}%
	\textcolor{black}{where $M_k$ ensures unit amplitude for the reflection coefficients.} \textcolor{black}{We prove in Appendix \ref{App:A} that, from an average perspective, the reflection channel can be approximated as
	\begin{equation}\label{Eq:H-app}
		{\bf H}\approx{{\bf{A}}_{{\rm b,e}}}{{\bf{\Xi }}_{\rm e}}{\bf{A}}_{{\rm u,e}}^H,
	\end{equation}
where ${ {{\bf{\Xi }}_{\rm e}} = {\rm diag} ( {{\xi _{1,0,0}}, \ldots ,{\xi _{K,0,0}}} ) }$ with ${\xi _{k,0,0}} = {\rho _k}g_{k,0}^{{{\rm rb}}}g_{k,0}^{{{\rm ur}}}$ representing the \emph{enhanced} gain of the cascaded LoS path in the UE--RIS$_k$--BS channel.} Here, ${{\bf{A}}_{{\rm b,e}}} = [ {{{\bf{a}}_{\rm b}}( {\Theta _{1,0}^{{{\rm rb}},{\rm A}}} ), \ldots ,{{\bf{a}}_{\rm b}}( {\Theta _{K,0}^{{{\rm rb}},{\rm A}}} )} ]$ and ${{\bf{A}}_{{\rm u,e}}} = [ {{{\bf{a}}_{\rm u}}( {\Theta _{1,0}^{{{\rm ur}},{\rm D}}} ), \ldots ,{{\bf{a}}_{\rm u}}( {\Theta _{K,0}^{{{\rm ur}},{\rm D}}} )} ]$ are the corresponding array response matrices at the BS and UE, respectively \cite{CF-2}.

Customizing a sparse reflection channel like \eqref{Eq:H-app} is advantageous for both uplink and downlink channel estimation. It transforms the transmission design paradigm in RIS-assisted systems, shifting from channel estimation followed by RIS configuration to a joint estimation and configuration approach. The desired reflection phase design in \eqref{Eq:RIS-design} is straightforward to implement with minimal complexities. Since $ \Theta _{k,0}^{{{\rm rb}},{\rm D}}$ and $\Phi _{k,0}^{{{\rm rb}},{\rm D}}$ can be determined by the positions of the BS and RIS, the necessities for channel customization are the AoDs of the enhanced path in the UE--RISs channel. Given that the required CSI for RIS configuration is quite limited and the customized channel is sparse, the uplink channel estimation can be significantly simplified by only extracting parameters of the strongest path, which are the LoS paths in the UE--RISs channel. Moreover, the overhead of downlink channel estimation can be reduced with this pre-designed sparse channel.  In the following subsections, we introduce the proposed low-complexity channel estimation.

\subsection{Positioning-based Joint LoS Paths Estimation}\label{sec:4-2}
In the channel customization under consideration, where RISs-cascaded LoS paths are enhanced with the array gain of RISs, parameters of $K$ cascaded LoS paths can be used to approximately reconstruct the uplink reflection channel. In this context, to reduce the channel estimation complexity, we focus on extracting the parameters of cascaded LoS paths rather than those of all paths.

Since the AoA and AoD of LoS paths in the RISs--BS channel can be derived from the fixed locations of the BS and RISs, only the parameters of LoS paths in the UE--RISs channel need to be estimated. As demonstrated in \eqref{Eq:NOMP-UE}, the NOMP algorithm can sequentially extract the parameters of each path. However, it cannot identify whether these parameters belong to the LoS path. Given that the LoS path is determined by the location of the UE, we propose a positioning-based joint LoS paths estimation algorithm.

Assuming that $K_{\rm L}$ out of $K$ RISs, which are denoted by ${\mathcal K}_{\rm L}$ within the set ${\mathcal K}$, have obtained the parameters of the LoS path during the extraction process of the NOMP algorithm, we can use the estimated $\{ {\hat \theta _{k,0}^{{{\rm ur}},{\rm A}},\hat \phi _{k,0}^{{{\rm ur}},{\rm A}},\hat \theta _{k,0}^{{{\rm ur}},{\rm D}}} \}_{k \in {{\mathcal K}_{\rm L}}}$ to calculate the position of the UE. Specifically, the direction vector of the LoS path in the UE--RIS$_k$ for $k\in{\mathcal K}_{\rm L}$ channel can be expressed as
\begin{equation}
	{{\bf{t}}_k} = {\left[ {\sin \hat \phi _{k,0}^{{{\rm ur}},{\rm A}}\cos \hat \theta _{k,0}^{{{\rm ur}},{\rm A}},\, \sin \hat \phi _{k,0}^{{{\rm ur}},{\rm A}}\sin \hat \theta _{k,0}^{{{\rm ur}},{\rm A}},\, \cos \hat \phi _{k,0}^{{{\rm ur}},{\rm A}}} \right]^T}.
\end{equation}
As the distance from the UE to RIS$_k$ is $d^{\rm ur}_{k}$, the UE's position observed at RIS$_k$ can be expressed as ${\hat{\bf{e}}_{{\rm u},k}} = {{\bf{e}}_k}+{d^{\rm ur}_{k}}{{\bf{t}}_k}$. Consequently, the averaged position error is defined as
\begin{equation}\label{Eq:f-e-d}
	f\left( {{{\bf{e}}_{\rm u}},{\bf{d}}} \right) = \sum\nolimits_{k \in {{\mathcal K}_{\rm L}}} {{{\left\| {{\hat{\bf{e}}_{{\rm u},k}} - {{\bf{e}}_{\rm u}}} \right\|}^2}}/{{K_{\rm L}}},
\end{equation}
where ${\bf{d}} = {\left[ {{d^{\rm ur}_{1}}, \ldots ,{d^{\rm ur}_{{K_{\rm L}}}}} \right]^T}$ is the distance vector of LoS paths in the UE--RISs channel. Given the direction vectors and positions of RISs in ${\mathcal K}_{\rm L}$, the UE's position can be estimated by
\begin{equation}\label{Eq:f-min}
	\left\{ {{{{\hat{\bf e}}}_{\rm u}},{\hat{\bf d}}} \right\} = \mathop {\arg \min }\limits_{{{\bf{e}}_{\rm u}},{\bf{d}}} f\left( {{{\bf{e}}_{\rm u}},{\bf{d}}} \right).
\end{equation}
The alternating optimization algorithm is used to solve \eqref{Eq:f-min}. When ${{\bf{e}}_{\rm u}}$ is fixed, the optimal ${\bf{d}}$ that minimizes $f( {{{\bf{e}}_{\rm u}},{\bf{d}}} )$ can be obtained by setting ${{\partial f\left( {{{\bf{e}}_{\rm u}},{\bf{d}}} \right)}}/{{\partial {\bf{d}}}} = 0$, whose solution can be expressed as
\begin{equation}\label{Eq:d}
	{\bf{d}} = {\bf{T}}{{\bf{e}}_{\rm u}} - {{\bf{d}}_{0}},
\end{equation}
where ${\bf{T}} = {[ {{{\bf{t}}_1},{{\bf{t}}_2}, \ldots ,{{\bf{t}}_{{K_{\rm L}}}}} ]^T} \in {{\mathbb C}^{{K_{\rm L}} \times 3}}$ and ${{\bf{d}}_{0}} = {[ {{\bf{t}}_1^T{{\bf{e}}_1},{\bf{t}}_2^T{{\bf{e}}_2}, \ldots ,{\bf{t}}_{{K_{\rm L}}}^T{{\bf{e}}_{{K_{\rm L}}}}} ]^T} $. Based on \eqref{Eq:d}, we have
\begin{equation}\label{Eq:e-u}
	{{\bf{e}}_{\rm u}} = {{\bf{T}}_{\rm P}}\left( {{\bf{d}} + {{\bf{d}}_{0}}} \right),
\end{equation}
where ${{\bf{T}}_{\rm P}}={( {{{\bf{T}}^T}{\bf{T}}} )^{ - 1}}{{\bf{T}}^T}$ is the pseudo-inverse of $\bf T$. \textcolor{black}{To ensure a unique solution for ${\bf e}_{\rm u}$, it is necessary that ${{\rm rank}({\bf T})\ge 3}$, implying that at least $K_{\rm L}=3$ RISs should be selected for positioning. Since ${\bf e}_{\rm u}$ can be expressed in terms of ${\bf d}$, \eqref{Eq:f-e-d} can be reformulated as}
\begin{equation}
	\begin{aligned}
		f\left( {\bf{d}} \right)=\frac{1}{{{K_{\rm L}}}}\sum\nolimits_{k \in {{\mathcal K}_{\rm L}}} &\left( \hat{\bf{e}}_{{\rm u},k}^T{\hat{\bf{e}}_{{\rm u},k}} - 2{\hat{\bf{e}}_{{\rm u},k}}^T{{\bf{T}}_{\rm P}}\left( {{\bf{d}} + {{\bf{d}}_{0}}} \right)\right.\\
		 +&\left. {{\left( {{\bf{d}} + {{\bf{d}}_{0}}} \right)}^T}{\bf{T}}_{\rm P}^H{{\bf{T}}_{\rm P}}\left( {{\bf{d}} + {{\bf{d}}_{0}}} \right) \right) .
	\end{aligned}
\end{equation}
The derivative of $f( {\bf{d}} )$ with respect to $\bf d$ is expressed as
\begin{equation}
	\begin{aligned}
		&\frac{{df( {\bf{d}} )}}{{d{\bf{d}}}}=\frac{2}{{{K_{\rm L}}}}\left( {{K_{\rm L}}{\bf{T}}_{\rm P}^T{{\bf{T}}_{\rm P}} + {\bf{I}} - 2{\bf{T}}{{\bf{T}}_{\rm P}}} \right){\bf{d}} \\
		+ \frac{2}{{{K_{\rm L}}}}&{\left( {{K_{\rm L}}{\bf{T}}_{\rm P}^T{{\bf{T}}_{\rm P}} + {\bf{I}} - {\bf{T}}{{\bf{T}}_{\rm P}}} \right)}{{\bf{d}}_{0}} - \frac{2}{{{K_{\rm L}}}}{\bf{T}}_{\rm P}^T\sum\limits_{k \in {{\mathcal K}_{\rm L}}} {{{\bf{e}}_k}} .
	\end{aligned}
\end{equation}
Setting ${{df( {\bf{d}} )}}/{{d{\bf{d}}}} = 0$, the optimal distance vector is estimated as
\begin{equation}
	\begin{aligned}
			&{{\bf{d}}^{\star}} = {\left( {{K_{\rm L}}{\bf{T}}_{\rm P}^T{{\bf{T}}_{\rm P}} + {\bf{I}} - 2{\bf{T}}{{\bf{T}}_{\rm P}}} \right)^{ - 1}}\\
			&\times\left( {{\bf{T}}_{\rm P}^T\sum\nolimits_{k \in {K_{\rm L}}} {{{\bf{e}}_k}}  - \left( {{K_{\rm L}}{\bf{T}}_{\rm P}^T{{\bf{T}}_{\rm P}} + {\bf{I}} - {\bf{T}}{{\bf{T}}_{\rm P}}} \right){{\bf{d}}_{0}}} \right).
	\end{aligned}
\end{equation}
The estimation for the UE's position is then given by
\begin{equation}
	{{\bf{e}}_{{\rm u}}^{\star}} = {{\bf{T}}_{\rm P}}\left( {{{\bf{d}}^{\star}} + {{\bf{d}}_{0}}} \right).
\end{equation}
According to the geometric relationship of ${{\bf{e}}_{{\rm u}}^{\star}}$ and $\{ {{{\bf{e}}_k}} \}_{k = 1}^K$, the AoAs and AoD of the LoS paths corresponding to RISs in ${\mathcal K}_{\rm L}$ can be calculated, denoted as $\{ {\tilde \theta _{k,0}^{{{\rm ur}},{\rm A}},\tilde \phi _{k,0}^{{{\rm ur}},{\rm A}},\tilde \theta _{k,0}^{{{\rm ur}},{\rm D}} \}_{k \in {\mathcal K}_{\rm L}}}$.

The proposed positioning algorithm requires the parameters of $K_{\rm L}$ LoS paths. To identify the LoS paths from the NOMP algorithm, we propose a positioning-based joint LoS paths estimation method, summarized in \textbf{Algorithm 1}. Specifically, whenever a path is detected from each UE--RIS link by the NOMP, $K_{\rm L}$ out of $K$ RISs are selected. This selection results in $C_K^{{K_{\rm L}}}= \frac{{K!}}{{{K_{\rm L}}!\left( {K - {K_{\rm L}}} \right)!}}$ combinations of $K_{\rm L}$ RISs. In each combination, each of the $K_{\rm L}$ RISs sequentially selects one path as the candidate for the LoS path. The averaged position error $f( {{{\bf{d}}^{\star}}} )$ is then calculated from these candidates. Assuming that $i$ paths are detected from each UE--RIS link, there will be ${i^{{K_{\rm L}}}}C_K^{{K_{\rm L}}}$ results of $f( {{{\bf{d}}^{\star}}} )$. The minimum of all possible $f( {{{\bf{d}}^{\star}}} )$ values is selected, and the corresponding ${{\bf{e}}_{{\rm u}}^{\star}}$ is regarded as the position of the UE. With the estimated position, the AoAs of the LoS path are calculated as $\{ {\tilde \theta _{k,0}^{{{\rm ur}},{\rm A}},\tilde \phi _{k,0}^{{{\rm ur}},{\rm A}}  \}_{k\in {\mathcal K}_{\rm L}}}$ and then compared with the NOMP-extracted AoAs $\{ \hat \theta _{k,l_k}^{{{\rm ur}},{\rm A}},\hat \phi _{k,l_k}^{{{\rm ur}},{\rm A}} \}_{{k\in {\mathcal K}},\, l_k\in\{1,2,\ldots,i\}}$.

When the AoAs of path $l_k^{\star}$ extracted by the NOMP are highly correlated with those calculated based on the estimated UE position, i.e., ${| {{\bf{a}}_{{\rm r},k}^H( {\hat \Theta _{k,l_{{k}}^{\star}}^{{{\rm ur}},{\rm A}},\hat \Phi _{k,l_{{k}}^{\star}}^{{{\rm ur}},{\rm A}}} ){{\bf{a}}_{{\rm r},k}}( {\tilde \Theta _{k,0}^{{{\rm ur}},{\rm A}},\tilde \Phi _{k,0}^{{{\rm ur}},{\rm A}}} )} |} \ge \tau $ holds for $\forall k\in{\mathcal K}$, $\{ {\hat \theta _{k,l_{{k}}^{\star}}^{{{\rm ur}},{\rm A}},\hat \phi _{k,l_{{k}}^{\star}}^{{{\rm ur}},{\rm A}},\hat \theta _{k,l_{{k}}^{\star}}^{{{\rm ur}},{\rm D}}} \}_{k \in {{\mathcal K}}}$ can be treated as the parameters of the LoS paths. In this case, we terminate the NOMP algorithm to avoid excessive computational complexity associated with estimating negligible paths. In typical mmWave systems, where the LoS path dominates the channel, our proposed algorithm is likely to detect the parameters of the LoS path during the first round of path extraction. Under these conditions, the minimum search complexity of our proposal is $KM{\eta ^3}( {{N_{\rm u}}+{N_{\rm b}}} )$, which is much smaller than \eqref{Eq:C-sim} of the NOMP. \textcolor{black}{Specifically, the NOMP method may require up to $KM{\eta ^3}( {({L^{{{\rm ur}}}-1)}{N_{\rm u}}+({L^{{{\rm rb}}}}-1){N_{\rm b}}} )$ additional coarse searches.} By using $\{ {\hat \theta _{k,l_{{k}}^{\star}}^{{{\rm ur}},{\rm A}},\hat \phi _{k,l_{{k}}^{\star}}^{{{\rm ur}},{\rm A}},\hat \theta _{k,l_{{k}}^{\star}}^{{{\rm ur}},{\rm D}}} \}_{k \in {{\mathcal K}}}$ to configure the reflection vector of the RISs, the uplink channel can be approximated by these LoS paths, as demonstrated in \eqref{Eq:H-app}.

\begin{algorithm}[htb]
	\caption{Positioning-based Joint LoS paths Estimation}
	\hspace*{0.02in} {\bf Input:} Received training signals in the uplink ${\bf{Y}}_k^{\rm b}$; stopping threshold $\tau $.\\
	\hspace*{0.02in} {\bf Output:} Parameters of LoS paths.
	\label{alg:Framwork}
	\begin{algorithmic}[1]
		
		\State Initialization: ${L^{{\rm ur}}_{\max }} = \mathop {\max }\nolimits_{k \in {\mathcal K}} {L_{k}^{\rm ur}}$ and $i=0$;
		\State{Sequentially select $K_{\rm L}$ out of $K$ RISs to construct $C_K^{K_{\rm L}}$ subsets $\{{\mathcal K}_{{\rm L},k}\}_{k=1}^{C_K^{K_{\rm L}}}$;}
		\While{$i \le {L^{{\rm ur}}_{\max }}$}
		 \For{$k = 1, \ldots ,K$}
		 \If{$i \le {L_{k}^{\rm ur}}$}
		 \State{Use NOMP to extract $\{ {\hat \theta _{k,i}^{{{\rm ur}},{\rm A}},\hat \phi _{k,i}^{{{\rm ur}},{\rm A}},\hat \theta _{k,i}^{{{\rm ur}},{\rm D}}} \}$ for the RIS$_k$;}
		 \EndIf
		 \EndFor
		 \For{$m = 1, \ldots ,C_K^{{K_{\rm L}}}$}
		 \State{Find the minimum position error for RISs set $m$: ${f_m}( {{{\bf{d}}_{{\rm est}}}} )=\mathop {{\rm min}}\limits_{n = 1, \ldots ,{i^{{K_{\rm L}}}}} {f_m}( {{{\bf{d}}_n}} )$;}
		 \EndFor
		 \State{Find the minimum position error of all RISs sets: $f( {{{\bf{d}}_{{\rm est}}}} ) = \mathop {{\rm min}}\limits_{m = 1, \ldots ,C_K^{{K_{\rm L}}}} {f_m}( {{{\bf{d}}_{{\rm est}}}} )$;}
		 \State{Derived the estimated UE position ${{\bf{e}}^{\star}_{{\rm u}}}$ corresponding to $f( {{{\bf{d}}_{{\rm est}}}} )$;}
		\State{Calculate the AoAs $\{ {\tilde \Theta _{k,0}^{{{\rm ur}},{\rm A}},\tilde \Phi _{k,0}^{{{\rm ur}},{\rm A}}} \}_{k\in {\mathcal K}_{\rm L}}$ according to ${{\bf{e}}^{\star}_{{\rm u}}}$ and $\{{\bf e}_k\}_{k\in{\mathcal K}_{\rm L}}$;}
		\State{Compare the calculated and extracted AoAs;}
		\If{${| {{\bf{a}}_{{\rm r},k}^H( {\hat \Theta _{k,l^{\star}_{{k}}}^{{{\rm ur}},{\rm A}},\hat \Phi _{k,l^{\star}_{{k}}}^{{{\rm ur}},{\rm A}}} ){{\bf{a}}_{{\rm r},k}}( {\tilde \Theta _{k,0}^{{{\rm ur}},{\rm A}},\tilde \Phi _{k,0}^{{{\rm ur}},{\rm A}}} )} |} \ge \tau $ holds for $\forall k\in{\mathcal K}$}
		\State Break;
		\Else
		 \State $i=i+1$
		\EndIf
		\EndWhile		
		\State  \Return $\{ {\hat \theta _{k,l^{\star}_k}^{{{\rm ur}},{\rm A}},\hat \phi _{k,l^{\star}_{{k}}}^{{{\rm ur}},{\rm A}},\hat \theta _{k,l^{\star}_{{k}}}^{{{\rm ur}},{\rm D}}} \}_{k \in {{\mathcal K}}}$;
	\end{algorithmic}
\end{algorithm}

\subsection{Low-overhead Downlink Channel Estimation}\label{sec:4-3}
For uplink channel estimation, a large number of pilots are required to train the complex, rich-scattering environment cascaded by RISs. By utilizing the prior CSI obtained in the uplink to configure RISs for channel customization, we can approximate the reflection channel with a few dominant paths. Based on this tailored channel, downlink channel estimation can be significantly simplified in terms of pilot overhead and estimation complexity.

Unlike uplink training, we omit the slow RIS reflection in the downlink training stage, reducing the number of pilots from $K_{\rm S}N_{\rm u}$ to $N_{\rm b}$, as the dominant components of the RIS-assisted channel are confirmed by the configuration given in \eqref{Eq:RIS-design}. The BS transmits downlink pilots ${{\bf{S}}_{\rm b}} = [ {{{\bf{s}}^{\rm b}_{1}},{{\bf{s}}^{\rm b}_{2}}, \ldots ,{{\bf{s}}^{\rm b}_{{N_{\rm b}}}}} ] \in {{\mathbb C}^{{N_{\rm b}} \times {N_{\rm b}}}}$ over $N_{\rm b}$ symbols. The transmit pilot vectors are mutually orthogonal and satisfy ${{\bf{S}}_{\rm b}}{\bf{S}}_{\rm b}^H = P_{\rm b}{\bf{I}}$. For the $q$-th pilot, $K_{\rm F}$ fast RIS reflections are performed to separate channels cascaded by different RISs. Specifically, the $v$-th fast RIS reflection of RIS$_k$ is given by ${f_{k,v}}\times {\rm diag}( {{\hat{\bm{\gamma }}_k}} )$, where $\hat{\bm \gamma}_k$ is determined by \eqref{Eq:RIS-design} with estimated AoAs. The UE's received signal for the $v$-th fast reflection in the $q$-th pilot is given by
\begin{equation}
	{{\bf{r}}^{\rm u}_{q,v}} = {{\bf{H}}^H}( v ){{\bf{s}}^{\rm b}_{q}} + {{\bf{n}}_{q,v}}.
\end{equation}
Similarly to the uplink training, after collecting signals of $K_{\rm K}$ fast reflections as ${{\bf{R}}^{\rm u}_{q}} = [ {{{\bf{r}}^{\rm u}_{q,1}}, \ldots ,{{\bf{r}}^{\rm u}_{q,{K_{\rm F}}}}} ] \in {{\mathbb C}^{{N_{\rm u}} \times {K_{\rm F}}}}$, we  multiply ${{\bf{R}}^{\rm u}_{q}}$ with the fast reflection matrix $\bf F$ to obtain
\begin{equation}
	{{\bf{R}}^{\rm u}_{q}}{\bf{F}} \triangleq \left[ {{{\bf{y}}^{\rm u}_{0,q}},{{\bf{y}}^{\rm u}_{1,q}}, \ldots ,{{\bf{y}}^{\rm u}_{K,q}}} \right] .
\end{equation}
Thanks to the orthogonality of ${\bf F}$, the signal components of the direct link and the reflected links are completely separated. For instance, the signal component delivered by the RIS$_k$ is given by
\begin{equation}
	{{\bf{y}}^{\rm u}_{k,q}} = {K_{\rm F}}{\rho _k}{\bf{H}}_{{{\rm ur}},k}^H{\rm diag}\left( {{\hat{\bm \gamma }}_k^H} \right){\bf{H}}_{{{\rm rb}},k}^H{{\bf{s}}^{\rm b}_{q}} + {{\bf{N}}_q}{{\bf{f}}_k}.
\end{equation}
Utilizing the channel approximation given in \eqref{Eq:H-app}, ${{\bf{y}}^{\rm u}_{k,q}}$ can be rewritten as
\begin{equation}
	{{\bf{y}}^{\rm u}_{k,q}} \approx {K_{\rm F}}{ \xi} _{k,0,0}^{{\rm i},*}{{\bf{a}}_{\rm u}}{\left( {\Theta _{k,0}^{{{\rm ur}},{\rm D}}} \right)}{\bf{a}}_{\rm b}^H{\left( {\Theta _{k,0}^{{{\rm rb}},{\rm A}}} \right)}{{\bf{s}}^{\rm b}_{q}} + {{\bf{N}}_q}{{\bf{f}}_k},
\end{equation}
where ${{ \xi }^{{\rm i}}_{k,0,0}} = {\rho _k}g_{k,0}^{{{\rm rb}}}g_{k,0}^{{{\rm ur}}}{\bf{a}}_{{\rm r},k}^H( {\Theta _{k,0}^{{{\rm rb}},{\rm D}},\Phi _{k,0}^{{{\rm rb}},{\rm D}}} ){\rm diag}( {{{{\hat{\bm\gamma }}}_k}} )\times$ ${{\bf{a}}_{{\rm r},k}}( {\Theta _{k,0}^{{{\rm ur}},{\rm A}},\Phi _{k,0}^{{{\rm ur}},{\rm A}}} )$ is the \emph{impaired} gain of the cascaded LoS path, which is unknown to the user. Stacking signals for $N_{\rm b}$ training symbols, we have
\begin{equation}
	\begin{aligned}
		&{{\bf{Y}}^{\rm u}_{k}} = \left[ {{{\bf{y}}^{\rm u}_{k,1}}, \ldots ,{{\bf{y}}^{\rm u}_{k,{N_{\rm b}}}}} \right]\\
		 &\approx {K_{\rm F}}{ \xi} _{k,0,0}^{{\rm i},*}{{\bf{a}}_{\rm u}}{\left( {\Theta _{k,0}^{{{\rm ur}},{\rm D}}} \right)}{\bf{a}}_{\rm b}^H{\left( {\Theta _{k,0}^{{{\rm rb}},{\rm A}}} \right)}{{\bf{S}}_{\rm b}} + {{\bf{N}}_{\rm u}}{\left( {{\bf{f}}_k^* \otimes {{\bf{I}}_{{N_{\rm b}} \times {N_{\rm b}}}}} \right)},
	\end{aligned}
\end{equation}
where ${{\bf{N}}_{\rm u}} \in {{\mathbb C}^{{N_{\rm u}} \times {K_{\rm F}}{N_{\rm b}}}}$ is the additive noise. Multiplying ${{\bf{Y}}^{\rm u}_{k}}$ with ${{{\bf{S}}_{\rm b}^H{{\bf{a}}_{\rm b}}( {\Theta _{k,0}^{{{\rm rb}},{\rm A}}} )}}/ (P {{{K_{\rm F}}}}) $, the training signal can be reduced to
\begin{equation}\label{Eq:y_CC-m}
	\begin{aligned}
		&{{\bf{y}}^{\rm u}_{k}} \triangleq {{{{\bf{Y}}^{\rm u}_{k}}{\bf{S}}_{\rm b}^H{{\bf{a}}_{\rm b}}{\left( {\Theta _{k,0}^{{{\rm rb}},{\rm A}}} \right)}}}/{({P_{\rm b}{K_{\rm F}}})} \\
		&\approx { \xi} _{k,0,0}^{{\rm i},*}{{\bf{a}}_{\rm u}}{\left( {\Theta _{k,0}^{{{\rm ur}},{\rm D}}} \right)} + {{\bf{N}}_{\rm u}}\left( {{\bf{f}}_k^* \otimes {{\bf{I}}_{{N_{\rm b}} \times {N_{\rm b}}}}} \right)\frac{{{\bf{S}}_{\rm b}^H{{\bf{a}}_{\rm b}}{\left( {\Theta _{k,0}^{{{\rm rb}},{\rm A}}} \right)}}}{{P_{\rm b}{K_{\rm F}}}}.
	\end{aligned}
\end{equation}
Thus, channel estimation for RIS-cascaded reflected links is simplified to multiple independent single-path extraction problems. The ML estimation for \eqref{Eq:y_CC-m} is given by
\begin{equation}\label{Eq:NOMP-DL}
	\begin{aligned}
		&\left( {{ {\check {\xi}}} _{k,0,0}^{{\rm i},*},\check{\Theta} _{k,0}^{{{\rm ur}},{\rm D}}} \right) = \mathop {\arg \min }\limits_{\left( {{ \xi} _{k,0,0}^{{\rm i},*},\Theta _{k,0}^{{{\rm ur}},{\rm D}}} \right)} {\left\| {{{\bf{y}}^{\rm u}_{k}} - { \xi} _{k,0,0}^{{\rm i},*}{{\bf{a}}_{\rm u}}{\left( {\Theta _{k,0}^{{{\rm ur}},{\rm D}}} \right)}} \right\|^2}\\
		&= \mathop {\arg \max }\limits_{\left( {{ \xi} _{k,0,0}^{{\rm i},*},\Theta _{k,0}^{{{\rm ur}},{\rm D}}} \right)} 2 {\rm Re} {\left\{ {{({\bf{y}}^{\rm u}_{k})}^H{ \xi} _{k,0,0}^{{\rm i},*}{{\bf{a}}_{\rm u}}{\left( {\Theta _{k,0}^{{{\rm ur}},{\rm D}}} \right)}} \right\}} - {\left| {{ \xi} _{k,0,0}^{{\rm i},*}} \right|^2}.
	\end{aligned}
\end{equation}
Given $\Theta _{k,0}^{{{\rm ur}},{\rm D}}$, the desired ${ {\check {\xi}}} _{k,0,0}^{{\rm i},*}$ that maximizes the object function can be derived as
\begin{equation}\label{Eq:check-xi}
	{ {\check {\xi}}} _{k,0,0}^{{\rm i},*} = {\bf{a}}_{\rm u}^H\left( {\Theta _{k,0}^{{{\rm ur}},{\rm D}}} \right){{\bf{y}}^{\rm u}_{k}}.
\end{equation}
Substituting ${ {\check {\xi}}} _{k,0,0}^{{\rm i},*}$ into \eqref{Eq:NOMP-DL} yields
\begin{equation}
	\check{\Theta} _{k,0}^{{{\rm ur}},{\rm D}} = \mathop {\arg \max }\limits_{\Theta _{k,0}^{{{\rm ur}},{\rm D}}} {\left| {{\bf{a}}_{\rm u}^H\left( {\Theta _{k,0}^{{{\rm ur}},{\rm D}}} \right){{\bf{y}}^{\rm u}_{k}}} \right|^2}.
\end{equation}
The coarse estimation of $\check{\Theta} _{k,0}^{{{\rm ur}},{\rm D}}$ can be found by searching over a predefined grid, and ${ {\check {\xi}}} _{k,0,0}^{{\rm i},*}$ is then updated according to \eqref{Eq:check-xi}. To improve the estimation accuracy, Newtonized refinement is applied for $\check{\Theta} _{k,0}^{{{\rm ur}},{\rm D}}$. When $\{ {\check \xi _{k,0,0}^{{\rm i},*},\check \Theta _{k,0}^{{{\rm ur}},{\rm D}}} \}_{k = 1}^K$ are obtained, the RIS-cascaded downlink reflection channel is then reconstructed as
\begin{equation}
	{{\check{\bf H}}}^H = {{\check{\bf A}}_{{\rm u,e}}}^H{\check{\bf \Xi }}_{\rm e}^H{{\bf A}}_{{\rm b,e}}^H,
\end{equation}
where ${{\check{\bf A}}_{{\rm u,e}}} = [ {{{\bf{a}}_{\rm u}}( {\check \Theta _{1,0}^{{{\rm ur}},{\rm D}}} ), \ldots ,{{\bf{a}}_{\rm u}}( {\check \Theta _{K,0}^{{{\rm ur}},{\rm D}}} )} ]$ and ${{\check{\bf \Xi }}_{\rm e}} = {\rm diag}( {{{\check  \xi }_{1,0,0}}^{\rm i}, \ldots ,{{\check \xi }_{K,0,0}^{\rm i}}} )$ are matrices reconstructed by parameters of \emph{enhanced} paths.

\subsection{Summary of the Proposed Low-complexity Channel Estimation}\label{sec:4-4}
\begin{figure}
	\centering
	\includegraphics[width=0.5\textwidth]{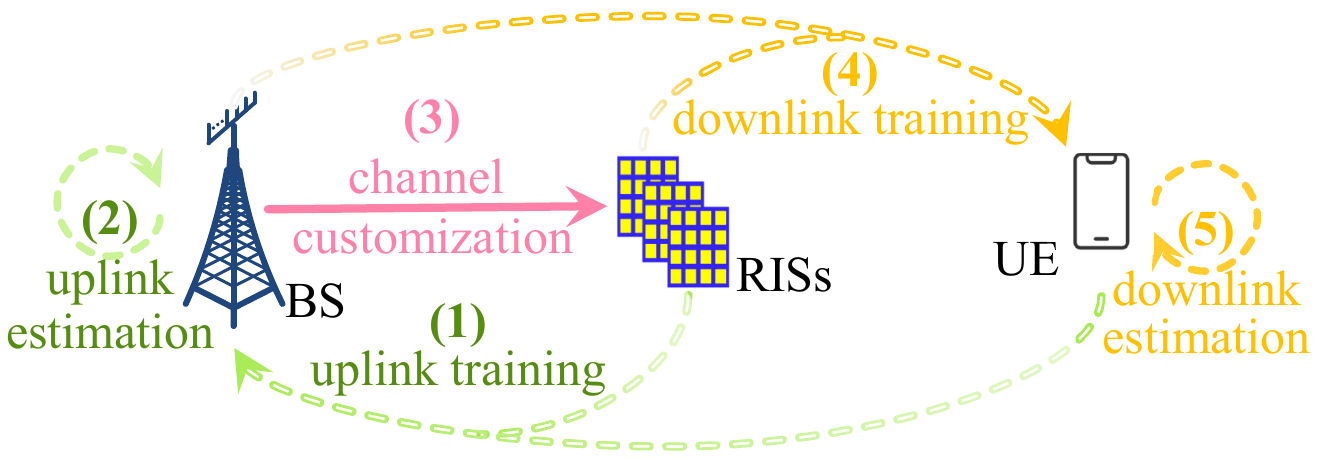}
	\caption{Overall flowchart of the proposed scheme.}
	\label{Fig.BD-CE}
\end{figure}

This subsection provides a summary of our approach to channel customization for efficient CSI acquisition. Illustrated in Fig. \ref{Fig.BD-CE}, the scheme unfolds in five stages\footnote{\textcolor{black}{The proposed scheme can be extended to multi-UE systems. Specifically, the methods in phases 1 and 2 can be directly applied when orthogonal pilots are used for UEs. However, customizing channels for multiple UEs in phase 3 introduces additional complexity, requiring considerations such as RIS resource allocation, RIS-UE association, and interference mitigation. Given the distinct channels being customized in multi-UE systems, the methodologies in phases 4 and 5 would need to be redesigned. Nevertheless, the approach using fast-varying RIS phase shifts for channel separation remains applicable and effective.}}:
\begin{itemize}
\item \emph{Phase 1:} Initially, with the BS and UE unaware of the rich-scattering environment, standard uplink training occurs. As Section \ref{sec:3-1} details, RISs are trained using slow-varying reflection vectors, with the UE transmitting pilots for each vector. The overlap of signals from various channel components at the BS, including direct and RIS-reflected links, complicates estimating individual links. To mitigate this, we employ fast-varying reflection phases across the RIS for each pilot, aiding in the separation of channel components.
\item \emph{Phase 2:} This stage focuses on uplink channel estimation for subsequent transmission design, deviating from the conventional CSI estimation then RIS configuration strategy. Our proposed joint approach targets only the dominant propagation paths, such as LoS paths, simplifying the process with a positioning-based algorithm for efficient LoS path identification.
\item \emph{Phase 3:} With LoS path parameters identified, we customize a sparse channel by adjusting RIS reflection vectors to highlight these dominant paths. This step allows the BS to approximate the channel matrix using the LoS path parameters, simplifying the channel to its essential components.
\item \emph{Phase 4:} Downlink training is streamlined in this phase, as the pre-customized sparse channel reduces the need for numerous pilots. Specifically, training now skips slow-varying RIS adjustments, focusing solely on fast-varying phases to further isolate channel components.
\item \emph{Phase 5:} The final phase simplifies downlink reflection channel estimation into discrete sub-problems, each concerning parameter detection for a single-path channel, facilitated by the preceding channel customization and training adjustments.

\end{itemize}

\section{Numerical Results}\label{sec:5}
In this section, we present numerical results to demonstrate the effectiveness of the proposed low-complexity CSI acquisition method. \textcolor{black}{Given the partial CSI estimation nature of our proposal, the performances of NOMP, a full CSI estimation algorithm, are evaluated as baselines.} The evaluations focus on the accuracy of parameter extraction, UE positioning, channel reconstruction, and the average SE in a mmWave MIMO system assisted by four RISs. Since the direct link between the BS and the UE is obstructed by a building, RISs are strategically placed in the LoS of the BS/UE to facilitate reliable data transmission. The system's carrier frequency is set at $26$ GHz. In the mmWave band, we set the Rician factor for the BS/RIS--UE channel at $10$ dB and for the RISs--BS channel at $30$ dB. The Cartesian coordinates for the BS and the four RISs are ${\bf e}_{\rm b}=[0,0,60]^T$, ${\bf e}_1=[86,-7,16]^T$, ${\bf e}_2=[71,-2,16]^T$, ${\bf e}_3=[71,-2,16]^T$, and ${\bf e}_4=[86,7,16]^T$, respectively. The UE is randomly located within a circular blind coverage area, centered at $[80,0,0]^T$ with a radius of $8$ meters. Unless stated otherwise, each RIS comprises $5\times 5$ elements, and the BS and UE have $16$ and $4$ antennas, respectively. We assume that the obstructing building has three concrete walls, each imposing a penetration loss of $26$ dB. The number of NLoS paths is set to 5 for the BS/RIS--UE channel and to 2 for the RISs--BS channel. During uplink training, DFT matrices are used for slow RIS reflection, indicating that $K_{\rm S}$ equals the number of elements in each RIS. The noise power is fixed at $-110$ dBm.

\subsection{Channel Separation and Sparsification}
In this subsection, we illustrate the RIS's channel customization capabilities, specifically focusing on channel separation and sparsification.
\begin{figure}
	\centering
	\includegraphics[width=0.5\textwidth]{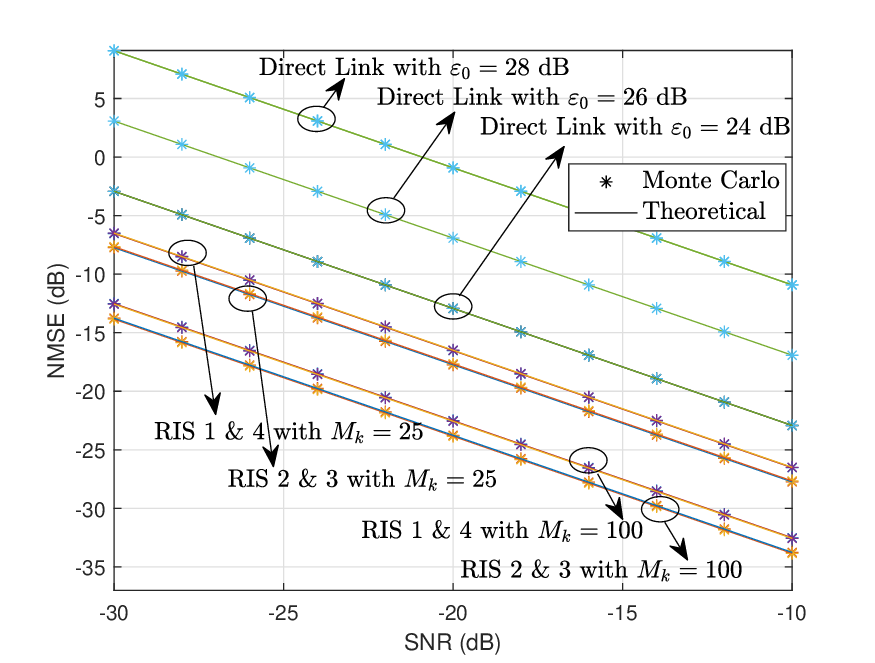}
	\caption{NMSE of the decoupled signal components.}
	\label{Fig.decouple}
\end{figure}
During the training phases, fast-varying RIS reflection is employed to distinguish signals from different channel components. The NMSE of the separated signal is defined as
\begin{equation}\label{Eq:decouple_MC}
	{{\left\| {{\bf{Y}}_k^{\rm b} - {K_{\rm F}}{\rho _k}\left( {\left( {{\bf{S}}_{\rm u}^T{\bf{H}}_{{{\rm ur}},k}^T} \right) \oplus {{\bf{H}}_{{{\rm rb}},k}}} \right){{\bf{\Gamma }}_k}} \right\|_F^2}}/{{\left\| {{\bf{Y}}_k^{\rm b}} \right\|_F^2}},
\end{equation}
where $k=0$ indicates the direct link, and $k>0$ refers to the reflection link cascaded by RIS$_k$. In Fig. \ref{Fig.decouple}, we compare the Monte Carlo results of \eqref{Eq:decouple_MC} with their theoretical counterpart, given by
\begin{equation}
	{{\left\| {{\rm vec}\left( {{{\bf{N}}_s}\left( {{{\bf{I}}_{{N_{\rm u}} \times {N_{\rm u}}}} \otimes {\bf{f}}_k^*} \right)} \right)} \right\|_F^2}}/{{\left\| {{\bf{Y}}_k^{\rm b}} \right\|_F^2}}.
\end{equation}
In 2,500 channel realizations, the Monte Carlo results align well with the theoretical NMSE, demonstrating that different link signal components can be effectively separated by the proposed fast-varying RIS reflection. The NMSE varies for different links due to the variations in corresponding larger-scale path loss. Specifically, the NMSE of the direct link deteriorates as the penetration loss increases. RIS$_1$ and RIS$_4$ (or RIS$_2$ and RIS$_3$), having similar deployment positions in the system, exhibit nearly identical performances. Additionally, a larger RIS scale can improve NMSE by collecting more channel power.

\begin{figure}
	\centering
	\subfigure[]{
		\label{Fig.ratio-M} 
		\includegraphics[width=0.5\textwidth]{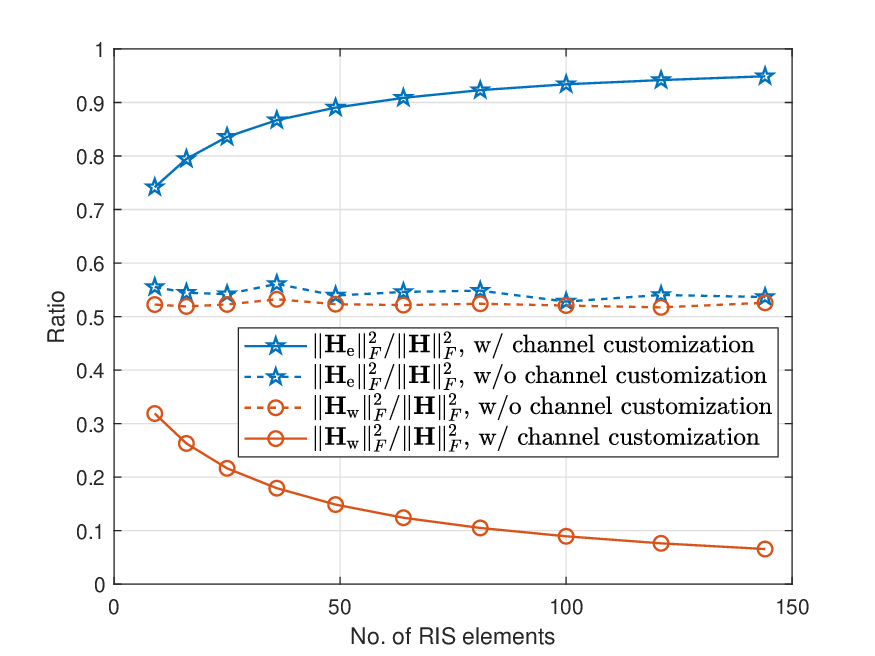}}
	\subfigure[]{
		\label{Fig.ratio-K} 
		\includegraphics[width=0.5\textwidth]{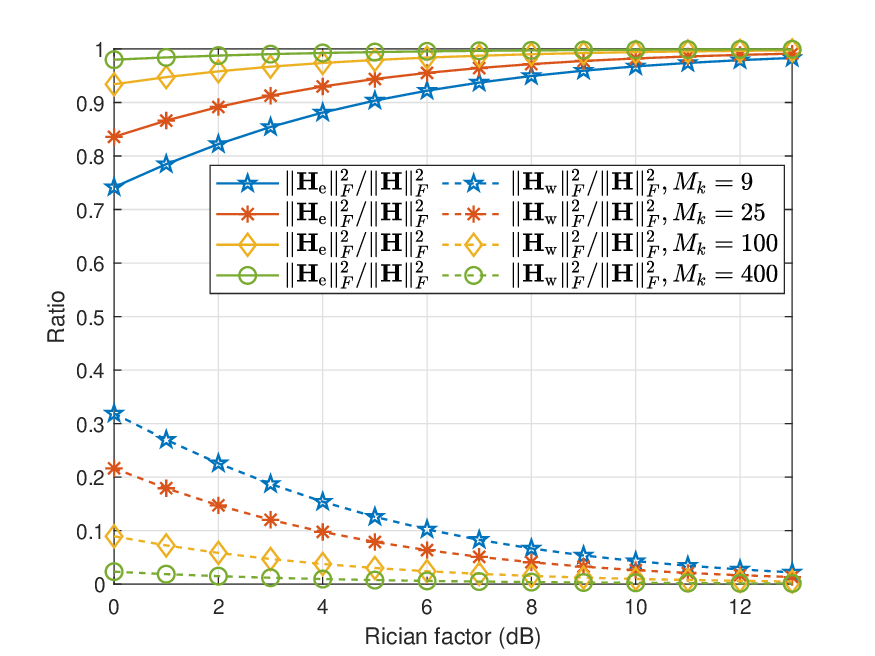}}
	\caption{Power ratio of channel components: (a) with and without channel customization for the Rician factor $\kappa_{{\rm ur}}=0$ dB; (b) impacts of $\kappa_{{\rm ur}}$ and $M_k$ for the customized channel.}
	\label{Fig.ratio} 
\end{figure}
Fig.~\ref{Fig.ratio} showcases the channel sparsification performance of the RIS configuration. In Fig. \ref{Fig.ratio-M}, where the Rician factor for the UE--RISs channel, $\kappa_{{{\rm ur}}}$, is set to $0$ dB, the power ratios of enhanced and weakened channel components, ${\bf H}_{\rm e}={{\bf{A}}_{{\rm b,e}}}{{\bf{\Xi }}_{\rm e}}{\bf{A}}_{{\rm u,e}}^H$ and ${\bf H}_{\rm w} = {\bf H}-{\bf H}_{\rm e}$, range between $0.5$ and $0.6$ when the channel is uncustomized (without RIS configuration). Implementing channel customization to enhance selected paths results in these paths dominating the reflection channel, indicating effective channel streamlining. The dominance of the enhanced paths increases with the number of RIS elements. Fig. \ref{Fig.ratio-K} shows that in typical mmWave systems with a Rician factor greater than $0$ dB, the power ratio of the dominant channel component gradually nears 1, as LoS paths are preferentially enhanced. From Figs. \ref{Fig.ratio-M} and \ref{Fig.ratio-K}, it is evident that the number of RIS elements has a similar impact on the power ratio as the Rician factor.

\subsection{LoS Paths Extraction in the Uplink and Downlink}
This subsection presents the normalized mean error (NME) of the estimated directional parameters. The NME of ${X^{{\rm est}}}\in \{ {\hat \Theta _{0}^{{{\rm ur}},{\rm A}},\hat \Phi _{0}^{{{\rm ur}},{\rm A}}}, \check{\Theta} _{0}^{{{\rm ur}},{\rm D}}\}$ is defined as
\begin{equation}
{\mathbb E}\left\{{{\sum\nolimits_{k = 1}^K {\left| {X_k^{{\rm est}} - X_k^{{\rm real}}} \right|} }}/{({2\pi K })}\right\},
\end{equation}
where ${X_k^{{\rm est}}}\in \{ {\hat \Theta _{k,l_k}^{{{\rm ur}},{\rm A}},\hat \Phi _{k,l_{{k}}}^{{{\rm ur}},{\rm A}}} \}$ or ${X_k^{{\rm est}}}= \check{\Theta} _{k,0}^{{{\rm ur}},{\rm D}}$ represents the estimated value corresponding to RIS$_k$ in the uplink or downlink, and $X_k^{{\rm real}}$ is the actual value of the LoS path.

\begin{figure}
	\centering
	\includegraphics[width=0.5\textwidth]{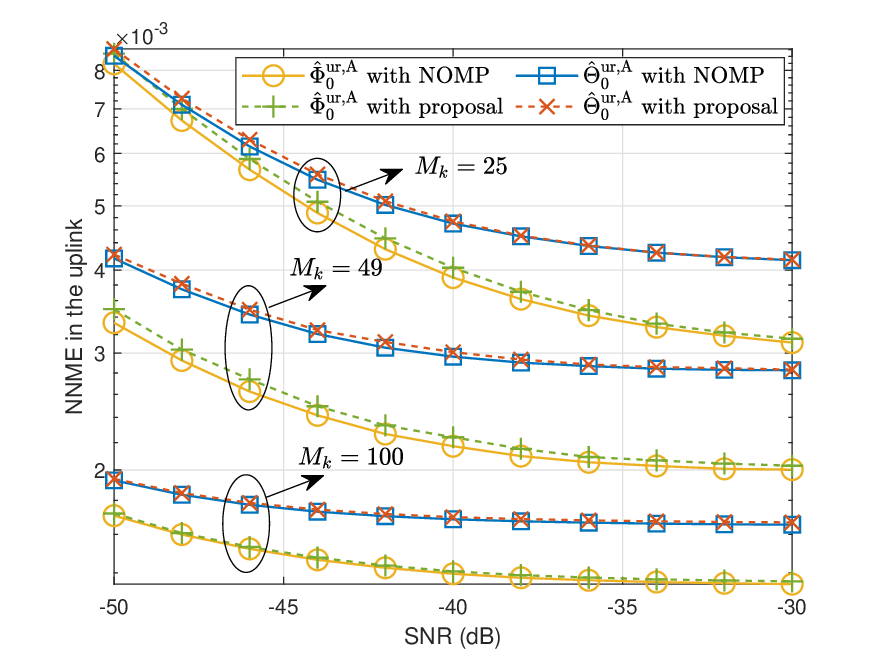}
	\caption{NME of parameters extracted in the uplink.}
	\label{Fig.UL-snr}
\end{figure}

Fig.~\ref{Fig.UL-snr} compares the NME of $\hat \Theta _{0}^{{{\rm ur}},{\rm A}}$ and $\hat \Phi _{0}^{{{\rm ur}},{\rm A}}$ as estimated by the NOMP and our proposed positioning-based algorithm. Given that $\hat\Theta _{k,l_k}^{\rm ur, A} = \pi \sin \hat\phi _{k,l_k}^{\rm ur, A}\sin \hat\theta _{k,l_k}^{\rm ur, A}$ and $\hat\Phi _{k,l_k}^{\rm ur, A} = \pi \cos \hat\phi _{k,l_k}^{\rm ur, A}$, the NME for $\hat \Theta _{0}^{{{\rm ur}},{\rm A}}$ is higher than that of $\hat \Phi _{0}^{{{\rm ur}},{\rm A}}$ due to the inclusion of errors in estimating both vertical and horizontal AoAs. As the SNR increases, both algorithms reach an NME plateau, which can be further lowered by enhancing the spatial resolution of beam training using a larger-scale RIS. Although our proposed algorithm exhibits slightly higher NME compared to the NOMP, it significantly reduces computational complexity. \textcolor{black}{For example, in cases where $M_{k}=25$, $49$, and $100$, the number of coarse searches for directional parameters is reduced by up to $9,600$, $18,816$, and $38,400$, respectively.}  Therefore, in subsequent simulations, we focus on results obtained using our proposed algorithm to better highlight our contributions.

\begin{figure}
	\centering
	\subfigure[NME versus $L^{\rm ur}_{k}$]{
		\label{Fig.UL-phi-path}
		\includegraphics[width=0.5\textwidth]{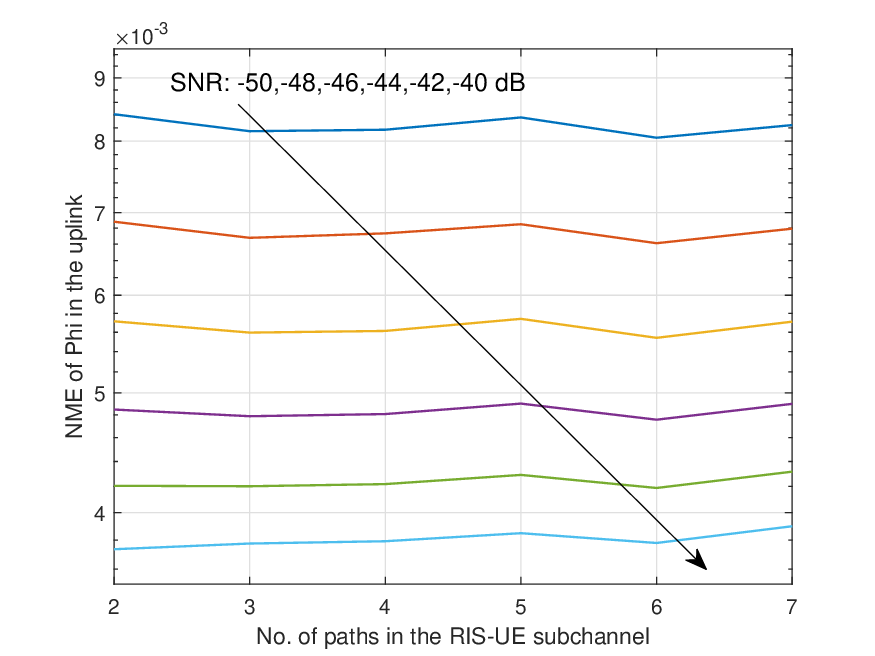}}\vskip -3pt
	\subfigure[NME versus $\kappa_{{\rm ur}}$]{
		\label{Fig.UL-phi-K}
		\includegraphics[width=0.5\textwidth]{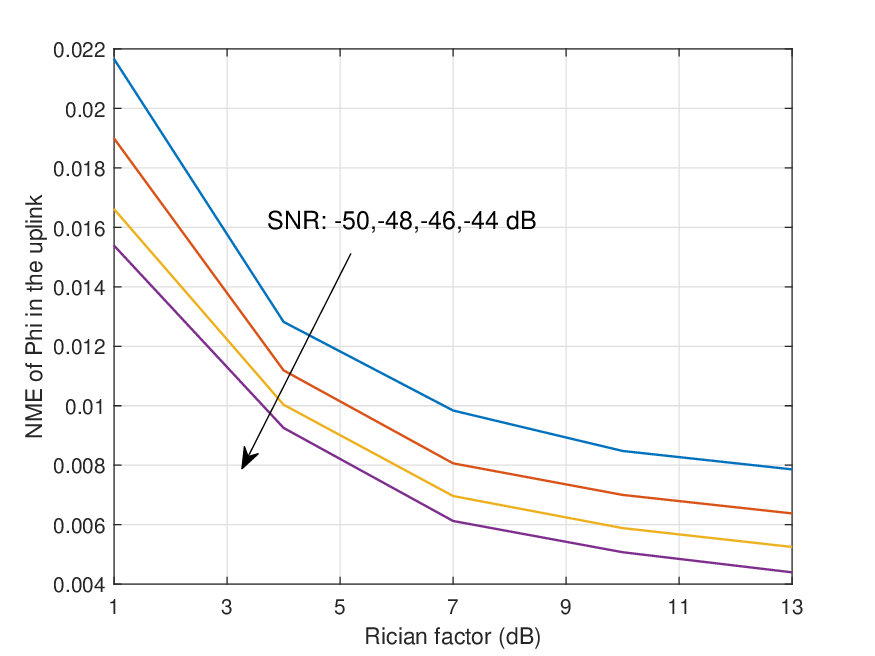}}
	\caption{NME of $\hat \Phi _{0}^{{{\rm ur}},{\rm A}}$ extracted in the uplink.}
	\label{Fig.UL-phi}
\end{figure}

Figs.~\ref{Fig.UL-phi-path} and \ref{Fig.UL-phi-K} display the NME of $\hat \Phi _{0}^{{{\rm ur}},{\rm A}}$ as a function of the number of paths and the Rician factor in the UE--RISs channel, respectively. Fig. \ref{Fig.UL-phi-path} demonstrates that our proposed method robustly extracts LoS path parameters even as the total number of paths increases. As depicted in Fig. \ref{Fig.UL-phi-K}, the estimation accuracy of LoS paths improves with a higher Rician factor.

\begin{figure}
 	\centering
 	\includegraphics[width=0.5\textwidth]{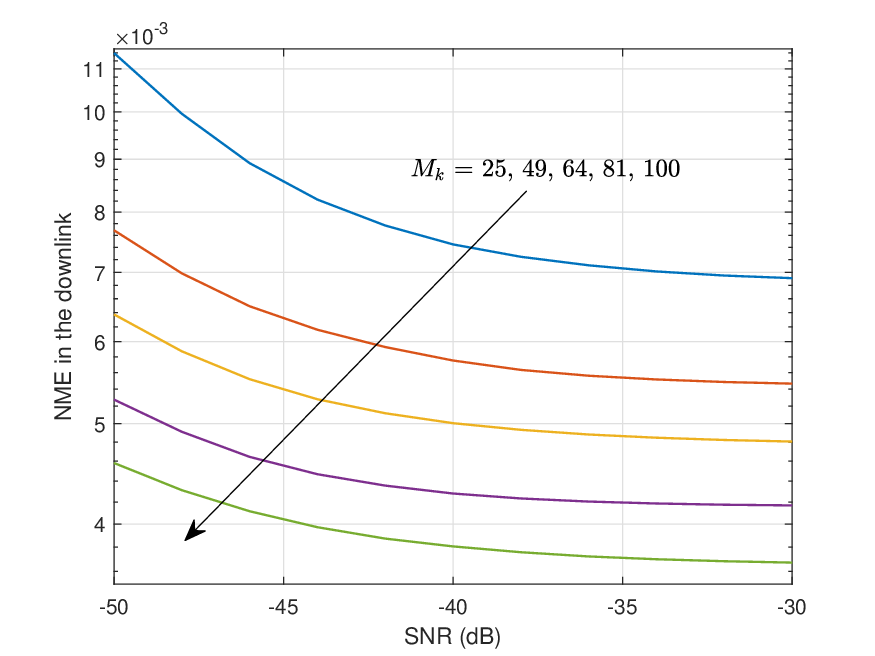}
 	\caption{NME of $\check{\Theta} _{0}^{{{\rm ur}},{\rm D}}$ extracted in the downlink.}
 	\label{Fig.DL-angle}
 \end{figure}

Building on the customized sparse channel concept introduced in Section \ref{sec:4-3}, we show that downlink channel estimation can be achieved with a simpler algorithm and fewer pilots. Fig. \ref{Fig.DL-angle} illustrates the NME of $\check{\Theta} _{0}^{{{\rm ur}},{\rm D}}$ extracted in the downlink. The NME in the downlink is only marginally inferior to that of the uplink, as shown in Fig. \ref{Fig.UL-snr}.

\subsection{Positioning}

\begin{figure}
	\centering
	\subfigure[Positioning error versus SNR]{
		\label{Fig.p-snr}
		\includegraphics[width=0.5\textwidth]{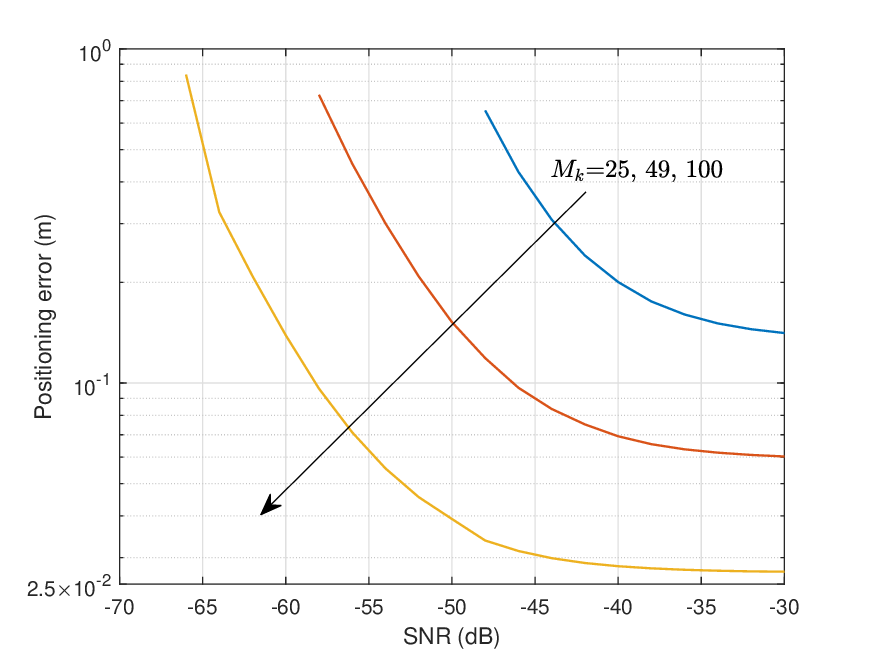}}\vskip -3pt
	\subfigure[Positioning error versus Rician factor $\kappa_{{{\rm ur}}}$]{
	\label{Fig.p-K}
	\includegraphics[width=0.5\textwidth]{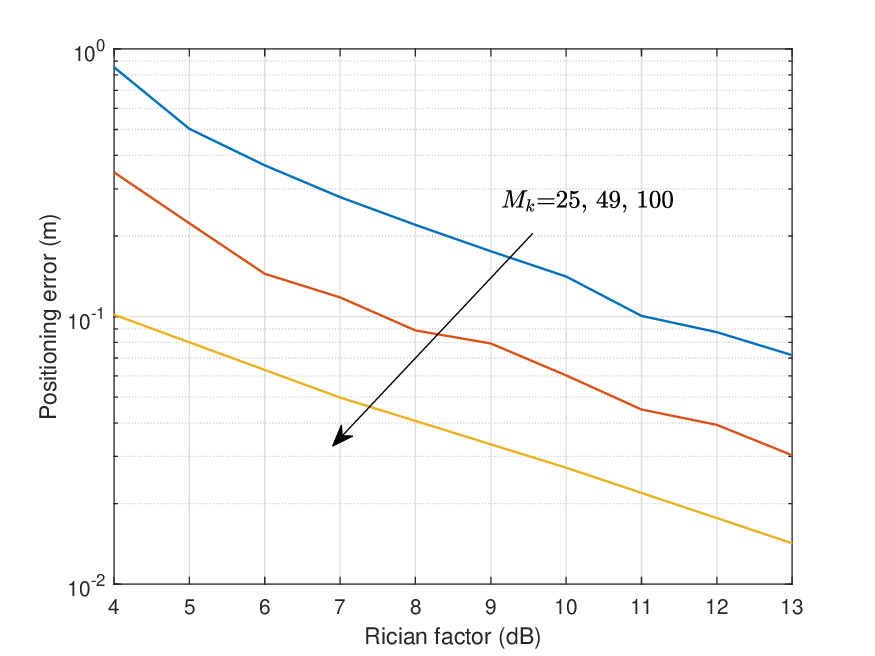}}
	\caption{Mean positioning error in the uplink.}
	\label{Fig.p}
\end{figure}

Using the AoAs extracted in the uplink, we assess the performance of the positioning algorithm as shown in Fig. \ref{Fig.p}. Under standard simulation conditions, Fig. \ref{Fig.p-snr} reveals that the positioning error reaches a plateau as the SNR increases. This trend is attributed to the angular estimation error bottleneck identified in Fig. \ref{Fig.UL-snr}, primarily caused by the interference from NLoS paths. In scenarios where the SNR is set to $-30$ dB, Fig. \ref{Fig.p-K} shows the impact of NLoS paths. As the Rician factor increases, the NLoS component diminishes, leading to a continuous decrease in positioning error.

\subsection{Channel Reconstruction}

In this subsection, we evaluate the performance of channel reconstruction in both the uplink and downlink, using the NMSE defined as $ {{\left\| {{\bf{H}} - {\bf{X}}} \right\|_F^2}}/{{\left\| {\bf{H}} \right\|_F^2}}$
where ${\bf X}={\hat{\bf H}}$ in the uplink and ${\bf X}={\check{\bf H}}$ in the downlink represent the reconstructed reflection channels.

\begin{figure}
	\centering
	\includegraphics[width=0.5\textwidth]{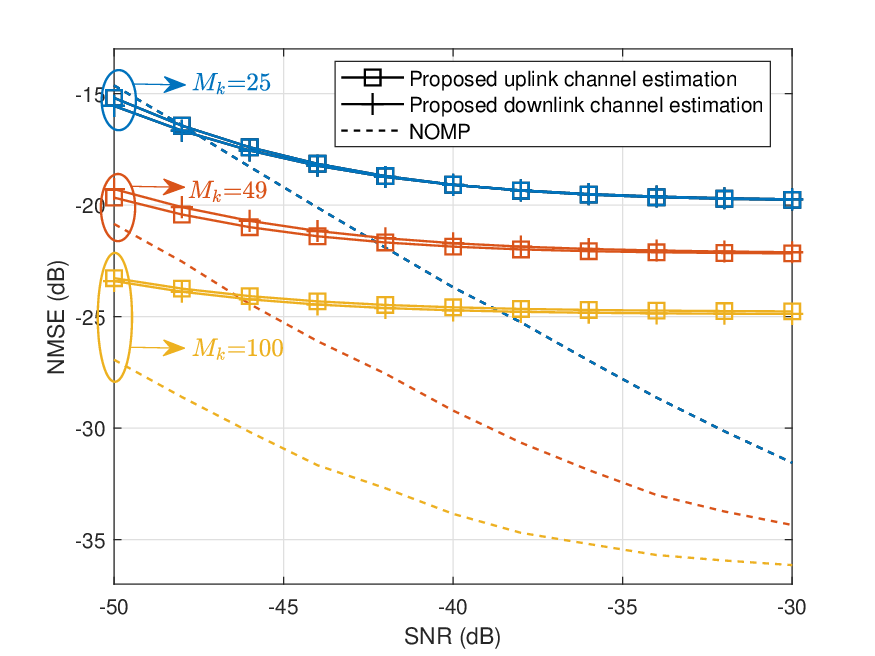}
	\caption{NMSE performance comparison of different schemes versus SNR with different RIS sizes.}
	\label{Fig.NMSE-M}
\end{figure}

Fig.~\ref{Fig.NMSE-M} compares the NMSE performances of various schemes against SNR for different RIS sizes. The NOMP algorithm, designed to extract path parameters regardless of path strength, consistently shows the best NMSE performance in channel reconstruction. In contrast, the NMSE performance of our proposed uplink channel estimation algorithm is more limited and does not continuously decrease with SNR as seen with the NOMP algorithm. This is due to the proposed algorithm focusing on estimating only the sparse paths enhanced by RISs. Despite having the same pilot overhead as the NOMP algorithm, our proposed algorithm benefits from lower computational complexity, which does not escalate with the increase in the number of channel propagation paths. In terms of downlink channel customization, the estimated channel primarily comprises enhanced LoS paths, leading to NMSE performances comparable to those in the uplink. Notably, both the pilot overhead and estimation complexity for the downlink are significantly lower than in the uplink, making the proposed downlink channel estimation particularly advantageous for UEs, which typically have much more limited signal processing capabilities compared to BS. Further analysis indicates that an increase in $M_k$ can reduce NMSE, aligning with the channel approximation presented in \eqref{Eq:H-app}.

\begin{figure}
	\centering
	\subfigure[NMSE versus SNR]{
	\label{Fig.NMSE-path}
	\includegraphics[width=0.5\textwidth]{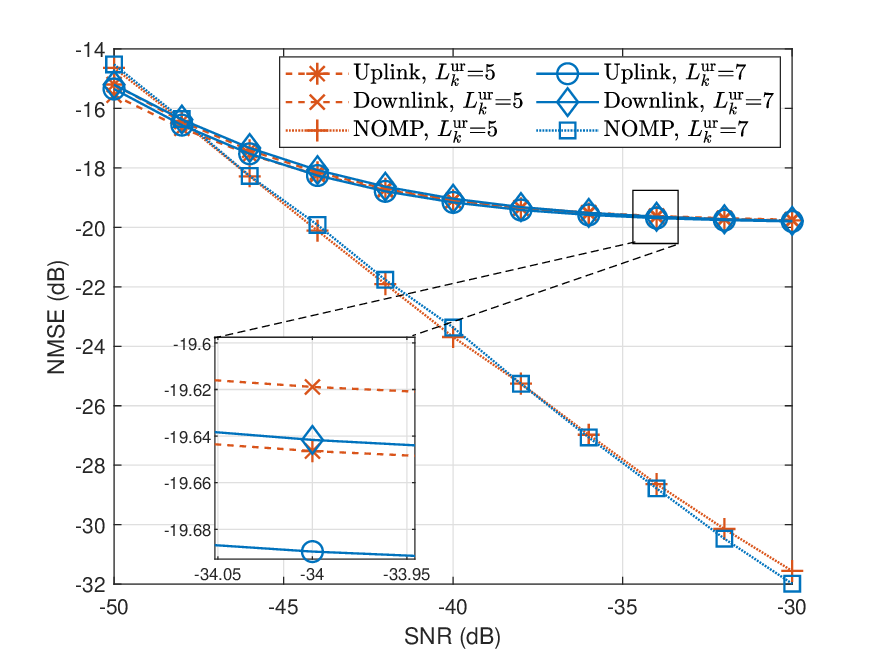}}\vskip -3pt
	\subfigure[\textcolor{black}{Complexity ratio between the NOMP and our proposal}]{
	\label{Fig.NMSE-path-ratio}
	\includegraphics[width=0.5\textwidth]{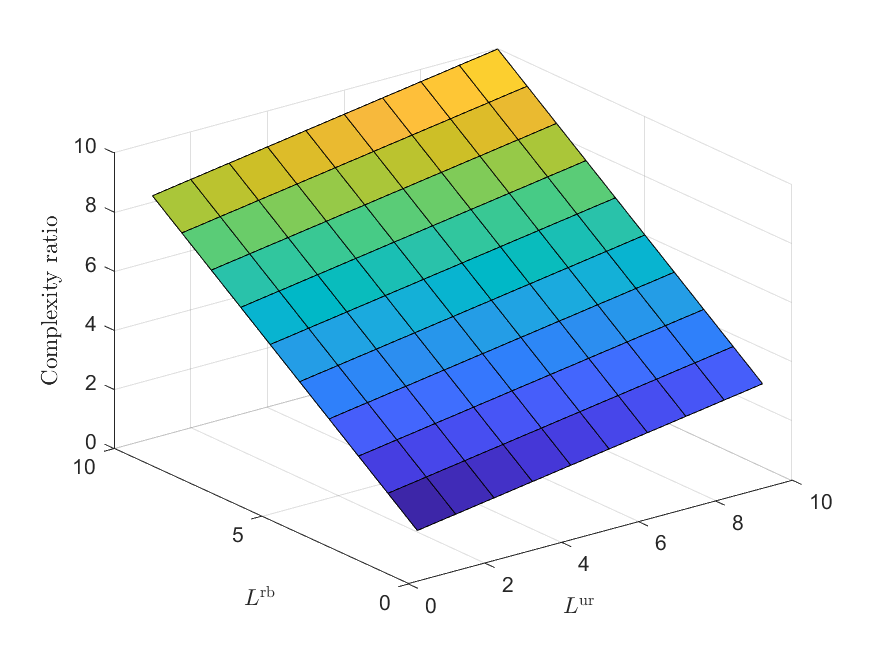}}
	\caption{\textcolor{black}{Impacts of the path number.}}
	\label{Fig.NMSE}
\end{figure}

\begin{figure}
	\centering
	\includegraphics[width=0.5\textwidth]{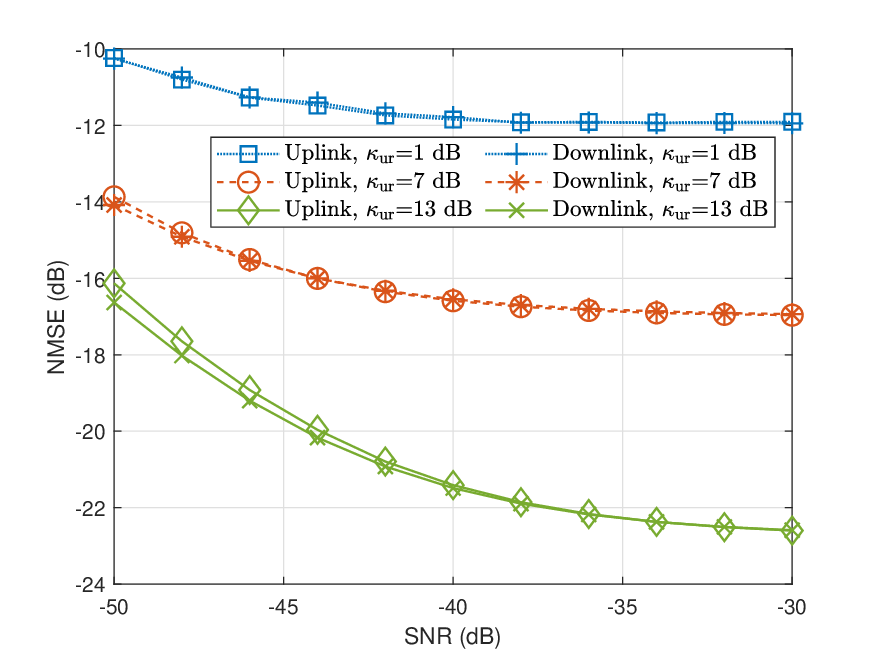}
	\caption{\textcolor{black}{Impacts of the Rician factor in the UE--RISs channel.}}
	\label{Fig.NMSE-K}
\end{figure}

\textcolor{black}{The impact of the number of paths and the Rician factor in the UE--RIS channels is explored in Figs. \ref{Fig.NMSE} and \ref{Fig.NMSE-K}, respectively}. Fig.~\ref{Fig.NMSE-path} shows that the NMSE performance is virtually unaffected by the number of paths, underscoring the robustness of the algorithms. \textcolor{black}{The search complexity ratio between the NOMP and our proposed method, $\frac{L^{\rm ur}N_{\rm u}+L^{\rm rb}N_{\rm b}}{N_{\rm u}+N_{\rm b}}$, is presented in Fig. \ref{Fig.NMSE-path-ratio}. In the default scenario, where the BS has more antennas than the UE, this ratio increases significantly with the number of paths in the BS-RIS channel. When the number of paths is high, our proposed method significantly outperforms the NOMP algorithm in terms of computational complexity. As LoS paths are preferentially enhanced and better channel approximations are achievable with higher Rician factors, the NMSE of the reconstructed channel is considerably reduced, as demonstrated in Fig. \ref{Fig.NMSE-K}.}

\subsection{Averaged SE}

\begin{figure}
	\centering
	\subfigure[averaged SE versus SNR]{
		\label{Fig.SE-M}
		\includegraphics[width=0.5\textwidth]{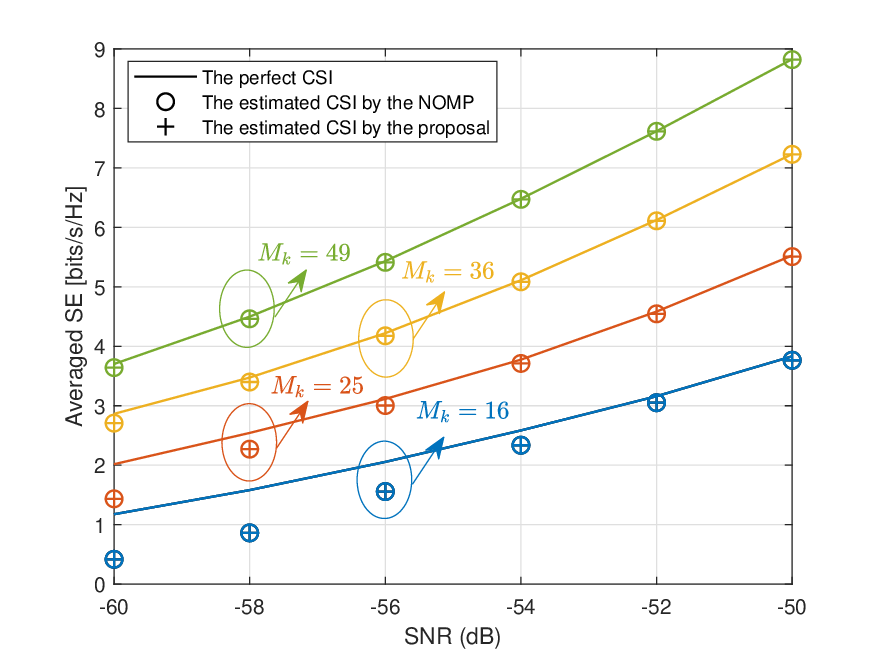}}\vskip -3pt
	\subfigure[averaged SE versus Rician factor $\kappa_{{{\rm ur}}}$]{
		\label{Fig.SE-K}
		\includegraphics[width=0.5\textwidth]{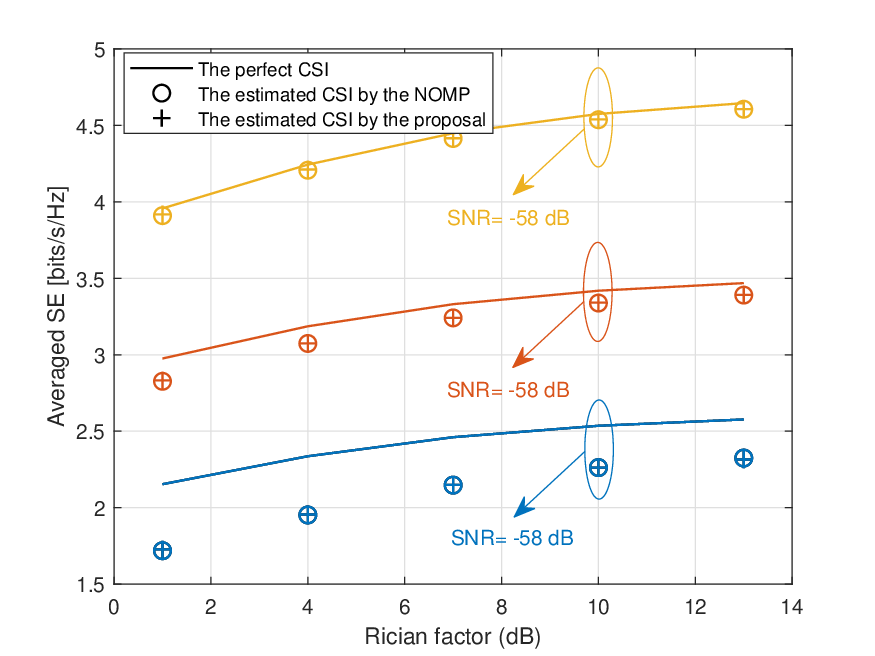}}
	\caption{averaged SE in the downlink with different CSI.}	
	\label{Fig.SE}
\end{figure}

In this subsection, we utilize the CSI obtained from both the NOMP and our proposed method to design the SVD transceiver. Fig. \ref{Fig.SE} illustrates the average of the SE defined in \eqref{Eq:R}. As observed in Fig. \ref{Fig.SE-M}, an increase in both SNR and the number of RIS elements results in the averaged SE performances of both the NOMP algorithm and our proposal closely approximating that achieved with perfect CSI. Notably, even though the NMSE of the channel reconstructed by our proposed algorithm is higher than that of the NOMP algorithm, the averaged SE performances of the two are very similar. This similarity suggests that the NMSE performance loss in our proposal remains within an acceptable range. Additionally, Fig. \ref{Fig.SE-K} demonstrates that as the Rician factor increases, the gap in averaged SE between perfect and estimated CSI narrows significantly.



\section{Conclusion}\label{sec:6}
This study introduced a novel approach to CSI acquisition in multi-RIS-assisted systems, representing a significant departure from the conventional sequential process of channel estimation followed by RIS reflection configuration. By leveraging the channel customization capabilities of RISs, our method integrates these processes, effectively addressing the challenge of signal coupling from direct and RIS-reflected links at the BS. Our strategy employs fast-varying phases across the RIS for each pilot symbol, creating orthogonal channel components that enable the effective separation of signals from different links at the BS. In the context of a complex reflection channel with numerous propagation paths, we utilized partial CSI for RIS configuration and introduced a positioning-based joint LoS paths estimation method. This approach emphasizes dominant paths, resulting in an approximately sparse reflection channel and significantly reducing the computational complexity traditionally associated with uplink channel estimation, which heavily relies on extracting path parameters. By leveraging this streamlined channel configuration, our method also substantially reduces both the pilot overhead and the computational complexity required for downlink channel estimation. Numerical results demonstrate the efficacy of RIS in channel separation and sparsification. Moreover, the proposed CSI acquisition method effectively extracts LoS paths and improves UE positioning. While the slight increase in NMSE in the reconstructed reflection channel remains within acceptable margins, it indicates a balanced trade-off between system complexity and SE.

\appendices
\section{}\label{App:A}
\color{black}When the reflection vectors are designed according to \eqref{Eq:RIS-design}, the cascaded path gain can be calculated as 
\begin{equation}
	\begin{aligned}
		{\xi _{k,l,c}} =& {\rho _k}g_{k,l}^{{{\rm rb}}}g_{k,c}^{{{\rm ur}}}\frac{{\sin \left( {{\Delta _{{\rm h},k,l,c}}{M_{{\rm h},k}}} \right)\sin \left( {{\Delta _{{\rm v},k,l,c}}{M_{{\rm v},k}}} \right)}}{{{M_{{\rm h},k}}\sin \left( {{\Delta _{{\rm h},k,l,c}}} \right){M_{{\rm v},k}}\sin \left( {{\Delta _{{\rm v},k,l,c}}} \right)}}\\
		&\times{e^{j\left( {{\Delta _{{\rm h},k,l,c}}\left( {{M_{{\rm h},k}} - 1} \right) + {\Delta _{{\rm v},k,l,c}}\left( {{M_{{\rm v},k}} - 1} \right)} \right)}},
	\end{aligned}
\end{equation}
where
\begin{equation}
	\left\{ \begin{aligned}
		{\Delta _{{\rm v},k,l,c}} &= \frac{{\Phi _{k,c}^{{{\rm ur}},{\rm A}} - \Phi _{k,0}^{{{\rm ur}},{\rm A}}}}{2} - \frac{{\Phi _{k,l}^{{{\rm rb}},{\rm D}} - \Phi _{k,0}^{{{\rm rb}},{\rm D}}}}{2}\\
		{\Delta _{{\rm h},k,l,c}} &= \frac{{\Theta _{k,c}^{{{\rm ur}},{\rm A}} - \Theta _{k,0}^{{{\rm ur}},{\rm A}}}}{2} - \frac{{\Theta _{k,l}^{{{\rm rb}},{\rm D}} - \Theta _{k,0}^{{{\rm rb}},{\rm D}}}}{2}
	\end{aligned} \right..
\end{equation}
Based on whether the LoS path is included or not, the averaged power of the cascaded path can be derived for four cases:
\begin{itemize}
	\item {\bf When $l=0$ and $c=0$:}
	\begin{equation}\label{Eq:App-00}
		{\mathbb E}\left\{{\left| {{\xi _{k,0,0}}} \right|^2}\right\} = {\left( {{\rho _k}g_{k,0}^{{{\rm rb}}}g_{k,0}^{{{\rm ur}}}} \right)^2} = \rho _k^2\frac{{M_k^2{N_{\rm b}}{N_{\rm u}}{\kappa _{{{\rm rb}}}}{\kappa _{{{\rm ur}}}}}}{{\left( {{\kappa _{{{\rm rb}}}} + 1} \right)\left( {{\kappa _{{{\rm ur}}}} + 1} \right)}}.
	\end{equation}
	\item {\bf When $l=0$ and $c>0$:}
	\begin{equation}\label{Eq:App-0m}
		\begin{aligned}
			&{\mathbb E}\left\{ {{{\left| {{\xi _{k,0,c}}} \right|}^2}} \right\} = \rho _k^2\frac{{M_k^2{N_{\rm b}}{N_{\rm u}}{\kappa _{{{\rm rb}}}}}{{\mathbb E}\left\{{\left| {\beta _{k,c}^{{{\rm ur}}}} \right|^2}\right\}}}{{\left( {{\kappa _{{{\rm rb}}}} + 1} \right)\left( {{\kappa _{{{\rm ur}}}} + 1} \right){L_{k}^{\rm ur}}}}\\
			&\times{\mathbb E}\left\{ {{{\left( {\frac{{\sin \left( {{\Delta _{{\rm h},k,0,c}}{M_{{\rm h},k}}} \right)\sin \left( {{\Delta _{{\rm v},k,0,c}}{M_{{\rm v},k}}} \right)}}{{{{M_{{\rm h},k}}}\sin \left( {{\Delta _{{\rm h},k,0,c}}} \right){M_{{\rm v},k}}\sin \left( {{\Delta _{{\rm v},k,0,c}}} \right)}}} \right)}^2}} \right\}.
		\end{aligned}
	\end{equation}
In \eqref{Eq:App-0m}, ${\Delta _{{\rm v},k,0,c}} = ( {\Phi _{k,c}^{{{\rm ur}},{\rm A}} - \Phi _{k,0}^{{{\rm ur}},{\rm A}}} )/2$. Assuming that $\Phi _{k,c}^{{{\rm ur}},{\rm A}} \sim [ { - \pi ,\pi } ]$, we have
\begin{equation}\label{Eq:App-inter-1}
	\begin{aligned}
		&{\mathbb E}\left\{ {\frac{{{{\sin }^2}\left( {{\Delta _{{\rm v},k,0,c}}{M_{{\rm v},k}}} \right)}}{{M_{{\rm v},k}^2{{\sin }^2}\left( {{\Delta _{{\rm v},k,0,c}}} \right)}}} \right\} \\
		&= \frac{1}{\pi }\int\limits_{-\pi {{/2 - }}\Phi _{k,0}^{{{\rm ur}},{\rm A}}/2}^{\pi {{/2 - }}\Phi _{k,0}^{{{\rm ur}},{\rm A}}/2} {\frac{{{{\sin }^2}\left( {{M_{{\rm v},k}}x} \right)}}{{M_{{\rm v},k}^2{{\sin }^2}\left( x \right)}}} {\rm D}x \\
		&= \frac{1}{\pi }\int\limits_{-\pi {{/2}}}^{\pi {{/2}}} {\frac{{{{\sin }^2}\left( {{M_{{\rm v},k}}x} \right)}}{{M_{{\rm v},k}^2{{\sin }^2}\left( x \right)}}} {\rm D}x,
	\end{aligned}
\end{equation}
where the second equation is derived because ${\frac{{{{\sin }^2}( {Nx} )}}{{{N^2}{{\sin }^2}( x )}}}$ has a periodicity of $\pi$. Utilizing the symmetry of integrals, \eqref{Eq:App-inter-1} can be further expressed as 
\begin{equation}\label{Eq:App-inter-2}
	\begin{aligned}
		&{\mathbb E}\left\{ {\frac{{{{\sin }^2}\left( {{\Delta _{{\rm v},k,0,c}}{M_{{\rm v},k}}} \right)}}{{M_{{\rm v},k}^2{{\sin }^2}\left( {{\Delta _{{\rm v},k,0,c}}} \right)}}} \right\} \\
		&= \frac{2}{{\pi M_{{\rm v},k}^2}}\int\limits_0^{\pi {{/2}}} {\frac{{{{\sin }^2}\left( {{M_{{\rm v},k}}x} \right)}}{{{{\sin }^2}\left( x \right)}}} {\rm D}x\\
		& = \frac{2}{{\pi M_{{\rm v},k}^2}}\frac{{{M_{{\rm v},k}}\pi }}{2}=\frac{1}{{{M_{{\rm v},k}}}},
	\end{aligned}
\end{equation}
where the second equation uses the result of \cite[Eq. (3.624)]{integrals}. Based on \eqref{Eq:App-inter-2}, we calculate \eqref{Eq:App-0m} as 
\begin{equation}\label{Eq:App-0m-1}
		{\mathbb E}\left\{ {{{\left| {{\xi _{k,0,c}}} \right|}^2}} \right\} = \rho _k^2\frac{{M_k{N_{\rm b}}{N_{\rm u}}{\kappa _{{{\rm rb}}}}}{{\mathbb E}\left\{{\left| {\beta _{k,c}^{{{\rm ur}}}} \right|^2}\right\}}}{{\left( {{\kappa _{{{\rm rb}}}} + 1} \right)\left( {{\kappa _{{{\rm ur}}}} + 1} \right){L_{k}^{\rm ur}}}}.
\end{equation}
Using \eqref{Eq:App-00} and ${{\mathbb E}\{{| {\beta _{k,c}^{{{\rm ur}}}} |^2}\}}=1$, \eqref{Eq:App-0m-1} can be expressed as 
\begin{equation}
		{\mathbb E}\left\{ {{{\left| {{\xi _{k,0,c}}} \right|}^2}} \right\}= {\mathbb E}\left\{{\left| {{\xi _{k,0,0}}} \right|^2}\right\}\frac{1}{{M_k}{L_{k}^{\rm ur}}{\kappa_{{\rm ur}}}}.
\end{equation}

	\item {\bf When $l>0$ and $c=0$:} ${\mathbb E}\{ {{{| {{\xi _{k,l,0}}} |}^2}} \}$ has a similar result to the previous case, given by
	\begin{equation}
		{\mathbb E}\left\{ {{{\left| {{\xi _{k,l,0}}} \right|}^2}} \right\}= {\mathbb E}\left\{{\left| {{\xi _{k,0,0}}} \right|^2}\right\}\frac{1}{{M_k}{L^{\rm rb}_k}{\kappa_{{\rm rb}}}}.
	\end{equation}

	\item {\bf When $l>0$ and $c>0$:}
	\begin{equation}
		\begin{aligned}
			&{\mathbb E} \left\{ {{{\left| {{\xi _{k,l,c}}} \right|}^2}} \right\} = \frac{\rho _k^2{M_k^2{N_{\rm b}}{N_{\rm u}}}}{{\left( {{\kappa _{{{\rm rb}}}} + 1} \right)\left( {{\kappa _{{{\rm ur}}}} + 1} \right){L_{k}^{\rm rb}}{L_{k}^{\rm ur}}}}\\
			&\times{\mathbb E}\left\{ {{{\left( {\frac{{\sin \left( {{\Delta _{{\rm h},k,l,c}}{M_{{\rm h},k}}} \right)\sin \left( {{\Delta _{{\rm v},k,l,c}}{M_{{\rm v},k}}} \right)}}{{{M_{{\rm h},k}}\sin \left( {{\Delta _{{\rm h},k,l,c}}} \right){M_{{\rm v},k}}\sin \left( {{\Delta _{{\rm v},k,l,c}}} \right)}}} \right)}^2}} \right\}.
		\end{aligned}
	\end{equation}
Assuming that ${\Phi _{k,c}^{{{\rm ur}},{\rm A}} \sim [ { - \pi ,\pi } ]}$ and ${\Phi _{k,l}^{{{\rm rb}},{\rm D}} \sim [ { - \pi ,\pi } ]}$, ${{{\Delta _{{\rm v},k,l,c}}}\sim[-\pi-{\delta _0},\pi-{\delta _0}]}$ can be derived with ${{\delta _0} = ( {\Phi _{k,0}^{{{\rm ur}},{\rm A}} - \Phi _{k,0}^{{{\rm rb}},{\rm D}}} )/2}$. Using the periodicity of ${\frac{{{{\sin }^2}( {Nx} )}}{{{N^2}{{\sin }^2}( x )}}}$ and following the results of previous cases, we have
\begin{equation}
	{\mathbb E}\left\{ {{{\left( {\frac{{\sin \left( {{\Delta _{{\rm h},k,l,c}}{M_{{\rm h},k}}} \right)\sin \left( {{\Delta _{{\rm v},k,l,c}}{M_{{\rm v},k}}} \right)}}{{{M_{{\rm h},k}}\sin \left( {{\Delta _{{\rm h},k,l,c}}} \right){M_{{\rm v},k}}\sin \left( {{\Delta _{{\rm v},k,l,c}}} \right)}}} \right)}^2}} \right\} = \frac{1}{{{M_k}}}.
\end{equation}
Thus, ${\mathbb E}	\{ {{{| {{\xi _{k,l,c}}} |}^2}} \}$ can be expressed as 
\begin{equation}
	{\mathbb E} \left\{ {{{\left| {{\xi _{k,l,c}}} \right|}^2}} \right\} ={\mathbb E}\left\{{\left| {{\xi _{k,0,0}}} \right|^2}\right\}\frac{1}{{{M_k}{L_{k}^{\rm rb}}{\kappa _{{{\rm rb}}}}{L_{k}^{\rm ur}}{\kappa _{{{\rm ur}}}}}}.
\end{equation}
\end{itemize}

The average power of all cascaded paths is given by
\begin{equation}
	{\mathbb E}\left\{ {\left\| {\bf{\Xi }} \right\|_F^2} \right\} = \sum\limits_{k = 1}^K {\mathbb E}{\left\{ {\left\| {{{\bf{\Xi }}_k}} \right\|_F^2} \right\}}.
\end{equation}
In the UE--RIS$_k$--BS channel, ${\mathbb E}{\{ {\| {{{\bf{\Xi }}_k}} \|_F^2} \}}$ can be expanded as 
\begin{equation}
	\begin{aligned}
		&{\mathbb E}\left\{ {\left\| {{{\bf{\Xi }}_k}} \right\|_F^2} \right\} = {\mathbb E}\left\{ {\sum\limits_{l = 0}^{{L_{k}^{\rm rb}}} {\sum\limits_{c = 0}^{{L_{k}^{\rm ur}}} {{{\left| {{\xi _{k,l,c}}} \right|}^2}} } } \right\} \\
		&= {\mathbb E}\left\{ {{{\left| {{\xi _{k,0,0}}} \right|}^2}} \right\} + \sum\limits_{c = 1}^{{L_{k}^{\rm ur}}} {\mathbb E}{\left\{ {{{\left| {{\xi _{k,0,c}}} \right|}^2}} \right\}}  + \sum\limits_{l = 1}^{{L_{k}^{\rm rb}}} {\mathbb E}{\left\{ {{{\left| {{\xi _{k,l,0}}} \right|}^2}} \right\}}\\
		&  + \sum\limits_{l = 1}^{{L_{k}^{\rm rb}}} {\sum\limits_{c = 1}^{{L_{k}^{\rm ur}}} {\mathbb E}{\left\{ {{{\left| {{\xi _{k,l,c}}} \right|}^2}} \right\}} } \\
		&={\mathbb E}{\left\{ {{{\left| {{\xi _{k,0,0}}} \right|}^2}} \right\}\left( {1 + \frac{1}{{{M_k}}}\left( {\frac{1}{{{\kappa _{{{\rm ur}}}}}} + \frac{1}{{{\kappa _{{{\rm rb}}}}}} + \frac{1}{{{\kappa _{{{\rm rb}}}}{\kappa _{{{\rm ur}}}}}}} \right)} \right)}.
	\end{aligned}
\end{equation}

For a sufficient larger $M_k$, which can be easily achieved since the RIS is typically equipped with a large number of elements, ${\mathbb E}\{ {\| {{{\bf{\Xi }}_k}} \|_F^2} \}\approx {\mathbb E}\{ {{{| {{\xi _{k,0,0}}} |}^2}} \}$ holds. Thus, the average power of all cascaded paths can be approximated as 
\begin{equation}
	{\mathbb E}\left\{ {\left\| {\bf{\Xi }} \right\|_F^2} \right\} \approx \sum\limits_{k = 1}^K {\mathbb E}\left\{ {{{\left| {{\xi _{k,0,0}}} \right|}^2}} \right\}={\mathbb E}\left\{ {\left\| {\bf{\Xi }}_{\rm e} \right\|_F^2} \right\},
\end{equation}
which implies that the power of all paths is dominated by the enhanced cascaded LoS path, and the reflection channel $\bf H$ is hardened to the dominant paths expressed by ${{\bf{A}}_{{\rm b,e}}}{{\bf{\Xi }}_{\rm e}}{\bf{A}}_{{\rm u,e}}^H$.
\color{black}



\begin{small}

\end{small}

\end{document}